\documentclass[useAMS]{mn2e}
\usepackage{graphicx,amssymb,amsmath}
\usepackage{hyperref,xcolor}
\usepackage{subfigure}
\usepackage{pdflscape}
\usepackage{mathrsfs}

\newcommand{\scs}{\scriptsize}

\title[He-enhanced giants in Omega Cen] {Omega Centauri: weak MgH-band in red giants directly trace the helium content }
\author[A. B. S. Reddy]
  { Arumalla B. S. Reddy$^{1}$\thanks{E-mail: balasudhakara.reddy@iiap.res.in }    \\
    $^1$Indian Institute of Astrophysics, II Block, Koramangala, Bangalore 560034, India }

\begin{document}

\date{Accepted 2020 April 16. Received 2020 April 12; in original form 2020 March 16}

\label{firstpage}
\pagerange{\pageref{firstpage}--\pageref{lastpage}} \pubyear{2020}

\maketitle

\begin{abstract}
High spectral resolution and high signal-to-noise ratio optical spectra of red giants in the globular cluster Omega Centauri are analysed for stellar parameters and chemical abundances of 15 elements including helium by either line equivalent widths or synthetic spectrum analyses. The simultaneous abundance analysis of MgH and Mg lines adopting theoretical photospheres and a combination of He/H$-$ratios proved to be the only powerful probe to evaluate helium abundances of red giants cooler than 4400 K, wherein otherwise helium line transitions (He {\scs I} 10830 and 5876 \AA) present for a direct spectral line analysis. The impact of helium-enhanced model photospheres on the resulting abundance ratios are smaller than 0.15 dex, in agreement with past studies. The first indirect spectroscopic helium abundances measured in this paper for the most metal-rich cluster members reveal the discovery of seven He-enhanced giants ($\Delta$$Y=+$0.15$\pm$0.04), the largest such sample found spectroscopically to date. The average metallicity of $-$0.79$\pm$0.06 dex and abundances for O, Na, Al, Si, Ca, Ti, Ni, Ba, and La are consistent with values found for the red giant branch (RGB-a) and subgiant branch (SGB-a) populations of Omega Centauri, suggesting an evolutionary connection among samples. 
The He-enhancement in giants is associated with larger $s$-process elemental abundances, which correlate with Al and anticorrelate with O. These results support the formation of He-enhanced, metal-rich population of Omega Centauri out of the interstellar medium enriched with the ejecta of fast rotating massive stars, binaries exploding as supernovae and asymptotic giant branch (AGB) stars.
\end{abstract}


\section{Introduction} 
A large number of spectroscopic and photometric studies of Omega Centauri (hereafter Omega Cen), the most massive known globular cluster (GC) of our Galaxy, have provided strong evidence of a colour spread on the colour-magnitude diagram (CMD) indicative of multiple sequences of stars along all evolutionary stages . Though the variation in metallicity is regarded as the first parameter resulting the width of an evolutionary sequence, variation in He content of stellar populations is introduced to account for the wide range of observed structure on CMDs (i.e., multiple main-sequences (MSs), subgiant branches (SGBs), red giant branches (RGBs) and an extended horizontal branch (HB)) of many globular clusters including the Omega Cen. Age and mass-loss of stars may also control the width of an evolutionary sequence on the CMD (Catelan 2009 for a review).

The complex chemical content of omega Cen was first revealed photometrically as an intrinsic width along the RGB in CMD (Cannon \& Stobie 1973), confirmed later as indicative of spread in metallicity, as noticed from RR Lyrae variables (Freeman \& Rodgers 1975). Later studies with wide field
photometry and spectroscopy have shown the co-existence of five RGBs of distinct metallicities ranging from about $-$2.0 dex to $-$0.6 dex (Lee et al. 1999; Sollima et al. 2005; Pancino et al. 2000; Johnson \& Pilachowski 2010), four to five SGB sequences of different ages and metallicities (Sollima et al. 2005; Villanova et al. 2007; Bellini et al. 2010), and at least three MSs (Anderson 1997; Bedin et al. 2004; Piotto et al. 2005; Sollima et al. 2005; Villanova et al. 2007). These multiple pathways on the cluster CMD are the result of different metallicities (Villanova et al. 2007; Marino et al. 2011; Johnson \& Pilachowski 2010; Simpson, Cottrell \& Worley 2012), helium abundances (Norris 2004; Piotto et al. 2005; Dupree et al. 2011) and CNO abundances (Marino et al. 2012; Simpson \& Cottrell 2013). 

Contrary to the standard stellar evolutionary model predictions, spectroscopy of MS stars in Omega Cen reveals that stars of intermediate metallicity ([Fe/H]$=-$\,1.37 dex) populate a bluer MS (bMS) with the most metal poor stars ([Fe/H]$=-$\,1.68 dex) on a redder MS (rMS; Piotto et al. 2005). To date, the only explanation offered to address the photometric and spectroscopic properties of MS stars is to assume a substantially enhanced helium abundance for the bMS stars ($Y=$\,0.35 $-$ 0.45) over the primordial He content of the rMS stars ($Y=$\,0.25; Bedin et al. 2004; Norris 2004; Piotto et al. 2005). A third, less populated MS (MS-a) found on the red side of rMS (Bedin et al. 2004) consists of the most metal-rich stars of Omega Cen ([Fe/H]$=-$\,0.6 dex) whose photometry based colours and metallicity are consistent with being populated with He-rich stars (Sollima et al. 2005; Bellini et al. 2010).

The most well studied peculiarities of Omega Cen are related to stellar metallicity distribution (Norris \& Da Costa 1995; Villanova et al. 2007; Johnson \& Pilachowski 2010; Simpson et al. 2012), variation in Na, Al and CNO elements along with anti-correlation between abundances of light elements (i.e., Na and O, Mg and Al) and a significant star-to-star scatter in $s$-process elements over the entire metallicity range (Stanford et al. 2007; Johnson \& Pilachowski 2010; Marino et al. 2011). 

The helium enhancement in second stellar generations is the only explanation for many observed features on CMDs (i.e., multiple MSs, RGBs and an extended HB) of all studied GCs (Marino et al. 2011 and references therein). Published spectroscopic abundance analyses have not focussed sharply on the possibility of measuring the helium content of GC stars, as these measurements are very difficult. The photospheric helium lines (He {\scs I} 5876 \AA) used for reliable He abundances are detectable in stars with effective temperatures between 8000 K and 11,500 K. Stars hotter than 11,500 K experience He sedimentation, which results He abundance in stellar photosphere not a representative of the original surface He content (Grundahl et al. 1999; Pace et al. 2009; Marino et al. 2014). In stars cooler than 8000 K, He lines are of chromospheric (infrared He {\scs I} 10830 \AA) rather than photospheric. A reliable estimate of He content of cool stars requires complex models that account for the chromospheric activity (Dupree, Strader \& Smith 2011). 

The helium abundances from He {\scs I} lines measured for a few GCs in the literature vary from very small amounts in NGC 6752 and NGC 6121 ($\Delta$$Y<$\,0.05; Villanova, Piotto \& Gratton 2009, Villanova et al. 2012) to larger values in Omega Cen and NGC 2808 ($Y=$\,0.35$-$0.40; Dupree et al. 2011; Pasquini et al. 2011; Marino et al. 2014). These analyses suggest that stars with higher He abundance are accompanied by enhancements of Na and Al indicative of high temperature H-burning including CNO, NeNa and MgAl cycles in previous stellar generations that lead to He-production (Gratton et al. 2001). The first acknowledgement of variation of He {\scs I} 10830 \AA\ line strengths (equivalent of abundance) among Omega Cen giants is provided by Dupree et al. (2011) for a sample of twelve giants covering a range of $-$1.87 dex to $-$1.16 dex in [Fe/H], 3 m\AA\ to 196 m\AA\ in He {\scs I} 10830 \AA\ line equivalent widths (EWs; line strengths) and effective temperatures between 4560 K and 4770 K. In spite of great interest in the astrophysical significance of helium enhancement in multiple stellar populations of Omega Cen, only two red giants out of the twelve in Dupree et al. (2011) have He abundances measured from the high-resolution (R$=$\,26,000 and 36,000) spectra (Dupree \& Avrett 2013).

This paper presents the first spectroscopic measurements of helium content for 13 cool red giants (10 giants out of 13 are cluster members) presumably populating the most metal-rich, anomalous red giant branch (named as RGB-a; Lee et al. 1999; Pancino et al. 2000) on the CMD of Omega Cen. To date, this is the largest of samples analysed for helium abundances in Omega Cen for which discrepant spectroscopic metallicities ranging from [Fe/H]$=-$\,1.1$\pm$0.2 dex (Villanova et al. 2007) to [Fe/H]$=-$\,0.60$\pm$0.15 dex (Pancino et al. 2002) exist in the literature. None of the giants on RGB-a have helium abundances measured directly from the spectra, but values ranging from $Y=$\,0.35 to $Y=$\,0.40 are suggested based on the compatibility of photometric colours and metallicities with helium enhanced theoretical isochrones of appropriate metallicity (Sollima et al. 2005; Piotto et al. 2005).

The helium abundance analysis technique adopted here for cool giants containing no photospheric helium lines in the optical spectra is based on the methodology advanced by Maeckle et al. (1975) for Arcturus: synthesis of atomic and hydride lines of an element in the observed spectra (i.e., C vs. CH and Mg vs. MgH) produce consistent values of [X/H] for an element X$=$ either C or Mg, if the adopted helium-to-hydrogen (He/H) ratio of the theoretical model atmosphere is compatible with that in the stellar photosphere. Differences larger than measurement errors in [X/H] abundance between atomic and hydride lines signify detectably different helium content of stars than that adopted in the standard stellar models. Such abundance differences are caused by the reduction in the dominant source of continuum opacity (i.e., H$^{-}$) in the stellar atmosphere due to He enrichment that enhances the atomic line strength (abundance) while reduces the line abundance of molecules bonded with hydrogen (Maeckle et al. 1975; Piotto et al 2005). This paper considers lines of only Mg and MgH to estimate the helium content of a star to avoid difficulties in continuum placement at the CH-band wavelengths (around 4300 \AA) due to heavy line blanketing and low signal-to-noise (S/N) ratio of the spectra.

This paper is organized as follows: Section 2 describes the spectral information and radial velocities of targets; Section 3 explains the stellar chemical composition analysis followed by validation of spectrum synthesis linelists for Mg and MgH lines; Section 4 summarizes the non-LTE Na, Mg, Al and Fe abundances; Section 5 compares chemical abundance results with literature studies, discusses the helium content and its association with abundances of O, Na, and Al in red giants; Section 6 discusses the chemical abundances and evolutionary connection of giants in this paper with the MS, SGB and RGB sequences on the CMD of Omega Cen. Finally, Section 8 provides concluding remarks.

\begin{table*}
\centering
\caption{The lis of Omega Cen red giants analysed in this paper.} 
\label{log_observations}
\begin{tabular}{lccccccc}   \hline
\multicolumn{1}{l}{Star}& \multicolumn{1}{c}{V} & \multicolumn{1}{c}{B-V} & \multicolumn{1}{c}{V-K$_{\rm s}$} & \multicolumn{1}{c}{J-K$_{\rm s}$} &
\multicolumn{1}{c}{RV (km s$^{-1}$)} & \multicolumn{1}{c}{RV (km s$^{-1}$) } & \multicolumn{1}{l}{Comment}  \\
\multicolumn{1}{c}{}& \multicolumn{1}{c}{(mag)}& \multicolumn{1}{c}{ } & \multicolumn{1}{c}{ }& \multicolumn{1}{c}{ } & 
\multicolumn{1}{c}{(This paper)} & \multicolumn{1}{c}{(Literature)} & \multicolumn{1}{l}{ }  \\
\hline

LEID~26067 & 13.675 & 1.295 & 3.128 & 0.828 & $+$245.4$\pm$0.2 & $+$244.4$\pm$0.9 & M \\
LEID~32149 & 13.568 & 1.302 & 3.496 & 0.888 & $+$235.4$\pm$0.2 & $+$234.3$\pm$1.2 & M \\
LEID~33164 & 13.260 & 1.200 & 3.380 & 0.855 & $+$245.3$\pm$0.4 & $+$245.7$\pm$0.1 & M \\
LEID~34029 & 12.107 & 1.491 & 3.388 & 0.916 & $+$207.1$\pm$0.7 & $+$208.2$\pm$0.6 & M \\
LEID~34180 & 13.030 & 1.481 & 3.700 & 1.010 & $+$236.9$\pm$0.6 & $+$234.3$\pm$1.3 & M \\
LEID~37024 & 12.740 & 1.710 & 4.269 & 1.104 & $+$227.0$\pm$0.7 & $+$226.1$\pm$0.3$^{\bf a}$ & M \\
LEID~41476 & 12.170 & 1.760 & 4.210 & 1.060 & $+$234.8$\pm$0.5 & $+$233.2$\pm$0.7 & M \\
LEID~42042 & 13.789 & 1.161 & 3.341 & 0.838 & $+$222.7$\pm$0.4 & $+$220.8$\pm$1.0 & M \\
LEID~48319 & 14.259 & 0.891 & $\ldots$ & $\ldots$ & $+$231.3$\pm$0.4 & $\ldots$ & M \\
LEID~54022 & 13.358 & 1.412 & 3.448 & 0.915 & $+$237.4$\pm$0.3 & $+$237.1$\pm$1.5 & M \\
LEID~32007 & 13.880 & 1.219 & $\ldots$ & $\ldots$ & $+$36.1$\pm$0.2 & $+$34.4$\pm$1.5 & NM \\
LEID~49026 & 13.475 & 1.212 & 3.243 & 0.829 & $+$50.3$\pm$0.3 & $+$50.8$\pm$1.6 & NM \\
LEID~78035 & 12.780 & 1.480 & 3.077 & 0.856 & $-$09.6$\pm$0.3 & $-$09.9$\pm$0.9 & NM \\

\hline
\end{tabular} 
\flushleft{ Note: ~M $=$ Member; ~NM $=$ non-member; ~$^{\bf a}$RV from Mayor et al. (1997) }
\end{table*}

\section{Observations and radial velocities} \label{obs_radial_velocities}
The spectra of RGB-a giants of Omega Cen analysed here were retrieved from the European Southern Observatory (ESO, Chile) public archive facility\footnote{\url{http://archive.eso.org/eso/eso_archive_main.html}}. 
Almost all the red giants populating the RGB-a on the cluster CMD are observed previously using the low (R$=$\,1600, $\lambda$$\sim$\,3840$-$4940 \AA) to medium (R$=$\,18,000, $\lambda$$\sim$\,6135$-$6365 and 6500$-$6800 \AA) resolution spectrographs and have chemical abundances measured for some of the elements by Simpson \& Cottrell (2013) and Johnson \& Pilachowski (2010). None of these spectra cover wavelength regions containing the lines of MgH (between 5130 $-$ 5160 \AA) and atomic Mg (5711, 6318.8 and 6319.2 \AA) to analyse them simultaneously for Mg abundances and then the He content. Moreover, the spectral line blending at low-resolution is a prime concern to determine reliable chemical composition of many elements. Therefore, the ESO archive was searched for the spectra of giants observed at the possible high spectral resolution and covering wavelengths of both MgH and Mg lines. This selection results 13 giants of which 10 are turned out to be cluster members with the remaining three giants positioned along the line-of-sight to the Omega Cen.

The spectra of selected giants were observed in 2001 April and 2007 January with the multi-object instrument FLAMES (Pasquini et al. 2000) equipped with UVES spectrograph at the 8-m VLT/UT2 telescope of the ESO. The fiber links to UVES and the CD3 cross-disperser were used to obtain high-resolution ($R=$\,42,310) spectra covering the wavelength range 4780$-$6805 \AA\ containing lines of many elements including the atomic Mg and molecular MgH. An automated FLAMES$-$UVES pipeline (Modigliani et al. 2004; Modigliani \& Larsen 2012) extracts the spectra to one-dimensional format and wavelength calibrates using Th-Ar lamp spectra as a reference. The radial velocity (RV) of each star was determined from the high-resolution spectrum by fitting central wavelengths of selected absorption lines (3 Fe{\scs I}, 3 Ca{\scs I}, 2 Na{\scs I}, Ni{\scs I} and Mn{\scs I}) with a Gaussian profile. The velocity measures from the individual lines have a dispersion of less than 1 km s$^{-1}$. The observed RVs were corrected for solar motion using the {\it rvcorrect} routine in {\scs IRAF}\footnote{IRAF is a general purpose software system for the reduction and analysis of astronomical data distributed by NOAO, which is operated by the Association of Universities for Research in Astronomy, Inc. under cooperative agreement with the National Science Foundation.}. The selected spectra were radial velocity shifted and normalized to unity.

The radial velocities for some of the stars in Omega Cen were measured previously in Mayor et al. (1997) and Reijns et al. (2006). The mean radial velocity of Omega Cen as shown in Reijns et al. (2006) has a distribution peaked at about $+$233 km s$^{-1}$ with a dispersion of about 30 km s$^{-1}$. The typical velocity dispersion of cluster members varies from 15 km s$^{-1}$ in the inner few arcmin to 6 km s$^{-1}$ at a cluster radius of about 25$\arcmin$. For the eleven targets, with the exception of LEID 37024 and LEID 48319, in common with Reijns et al., RV measurements in Table \ref{log_observations} are in excellent agreement with a mean difference of $+$0.8$\pm$1.1 km s$^{-1}$. Of the remaining two stars, LEID 48319 has no previous value of RV while the radial velocity of LEID 37024 agrees well with the value of $+$226.1$\pm$0.2 km s$^{-1}$ reported in Mayor et al. (1997). All but the three stars, namely LEID 32007, LEID 49026, and LEID 78035, listed in Table \ref{log_observations} have RVs consistent with that of cluster mean velocity confirming previous membership determination based on proper motions of stars by van Leeuwen et al. (2000).

\begin{table*}
\centering
\begin{minipage}{150mm}
\caption{Photometric and spectroscopic atmospheric parameters for open cluster members analysed in this study. }
\label{stellar_param}
\begin{tabular}{lcccccccccc}  \hline
\multicolumn{1}{l}{Star}& \multicolumn{3}{c}{T$^{\rm phot}_{\rm eff}$ (K)} &
\multicolumn{1}{c}{$\log g^{(B-V)}_{\rm phot}$}& \multicolumn{1}{c}{T$^{\rm spec}_{\rm eff}$}& 
\multicolumn{1}{c}{$\log g^{\rm spec}$}& \multicolumn{1}{c}{$\xi^{\rm spec}_{t}$}&
\multicolumn{2}{c}{$\log(L/L_\odot)$} \\  
\cline{2-4}
\cline{9-10}
\multicolumn{1}{c}{LEID}& \multicolumn{1}{c}{(B-V)}& (V-K)& (J-K)& 
\multicolumn{1}{c}{(cm s$^{-2}$)}& \multicolumn{1}{c}{(K)} & \multicolumn{1}{c}{(cm s$^{-2}$)}& 
\multicolumn{1}{c}{(km s$^{-1}$)}& \multicolumn{1}{c}{spec} & \multicolumn{1}{c}{phot} \\
\hline
     26067 &  4161 &  4302 &  4187 &  1.60 &  4250 &  1.40 & 1.43 & 2.41 &  2.17 \\
     32149 &  4149 &  4074 &  4049 &  1.55 &  4200 &  1.35 & 1.57 & 2.44 &  2.22 \\
     33164 &  4320 &  4140 &  4123 &  1.55 &  4250 &  1.30 & 1.80 & 2.51 &  2.28 \\
     34029 &  3877 &  4135 &  3989 &  0.72 &  4100 &  0.70 & 1.77 & 3.05 &  2.93 \\
     34180 &  3890 &  3969 &  3805 &  1.10 &  4000 &  1.00 & 1.67 & 2.70 &  2.56 \\
     37024 &  3616 &  3741 &  3644 &  0.56 &  3825 &  0.70 & 1.59 & 2.93 &  2.97 \\
     41476 &  3564 &  3761 &  3717 &  0.18 &  3800 &  0.30 & 1.74 & 3.32 &  3.32 \\
     42042 &  4390 &  4163 &  4163 &  1.81 &  4200 &  1.40 & 1.60 & 2.39 &  2.05 \\
     48319 &  4972 &$\ldots$&$\ldots$&2.33 &  4100 &  1.10 & 1.65 & 2.65 &  1.75 \\
     54022 &  3985 &  4100 &  3991 &  1.33 &  4100 &  1.20 & 1.62 & 2.55 &  2.37 \\
     32007 &  4287 &$\ldots$&$\ldots$&1.78 &  4350 &  1.95 & 1.30 & 1.90 &  2.05 \\
     49026 &  4299 &  4225 &  4184 &  1.63 &  4275 &  1.60 & 1.35 & 2.22 &  2.20 \\
     78035 &  3891 &  4338 &  4121 &  1.00 &  4225 &  1.30 & 1.65 & 2.50 &  2.65 \\
\hline
\end{tabular}
\end{minipage}
\end{table*}

The spectra of stars considered for abundance analysis in this paper have typical signal-to-noise ratio (S/N) of about 120 at wavelengths of the prominent MgH lines between 5130 $-$ 5160 \AA\ and S/N$>$150 at 5711 \AA\ and 6319 \AA\ regions containing the atomic Mg lines. Six giants with LEID numbers 26067, 32149, 33164, 41476, 42042, and 54022 have elemental abundances measured for C, N, O, Fe and Ba using the low resolution spectra (R$=$1600) covering the wavelength range 3840$-$4940 \AA (Simpson \& Cottrell 2013). Four stars with LEID numbers 34029, 34180, 37024, and 41476 were explored previously using medium-resolution spectra (R$=$\,18,000) for stellar parameters and chemical abundances of the elements O, Na, Si, Ca, Ti, Ni, Fe and La by Johnson \& Pilachowski (2010). The spectra (R$=$\,20,000$-$25,000) of five stars with LEID numbers 26067, 34029, 41476, 42042, and 54022 were analysed for abundances of elements Fe, O, Na, Ba and La by Marino et al. (2011). The stars with LEID numbers 32007, 48319, 49026, and 78035 have no previous record of chemical composition. This paper provides from the uniform abundance analysis of high-resolution spectra (R$=$\,42,310) the chemical abundances of elements O, Na, Mg, Al, Si, Ca, Ti, Cr, Fe, Ni, Y, Zr, Ba, La, Ce, and Sm including the first spectroscopic estimate of the He content for all giants via the synthesis of MgH and Mg lines in the observed spectra.

\section{Stellar parameters and chemical composition}
Following the standard spectroscopic abundance analysis technique, the stellar parameters and chemical abundances of stars are derived using either line equivalent widths or synthetic spectrum analyses of a combination of selected spectral lines of elements (linelist), a set of theoretical photospheres and a spectral analysis code. 

\subsection{Linelist}
The linelist containing the atomic line data was taken from Reddy, Giridhar \& Lambert (2012, 2015) and the line EWs were measured manually from the observed spectra using the {\it splot} task in {\scs IRAF} by either Gaussian profile fitting or direct integration of the observed absorption lines. The spectral linelist includes clean, unblended, relatively isolated and symmetric spectral lines of 16 elements, as noted earlier, and the MgH molecule. The molecular line data for MgH was extracted from Hinkle et al. (2013).

Stellar chemical abundances of most elements well represented by lines are measured from lines weaker than EW$=$130 m\AA, but relatively strong lines (EW$\sim$ 200 m\AA) were included for species represented by a few lines (for example, Ba {\scs II}). The final list of 200 absorption lines contains on average 80 Fe {\scs I} lines covering a range of 0.0$-$5.0 eV in line's lower excitation potential (LEP) and 15$-$130 m\AA\ in EWs, and 10 Fe {\scs II} lines spanning LEPs of about 2.9 to 3.9 eV and EWs from $\simeq$ 15 to 70 m\AA. The magnesium abundance is derived using three atomic lines and several clean MgH lines, as explained in the Section \ref{validation_mg_mgh}.

\subsection{Stellar parameters} \label{Stellar_parameters}
As the spectral line strength is affected by physical conditions and number density of absorbers in the photosphere, it is essential to predetermine the atmospheric parameters (e.g., effective temperature, T$_{\rm eff}$, and surface gravity, $\log~g$) for estimating stellar chemical abundances. Initial estimates of T$_{\rm eff}$ and $\log~g$ were obtained using the optical and 2MASS (Cutri et al. 2003) photometric colours (B-V), (V-K$_{\rm s}$) and (J-K$_{\rm s}$) and equations discussed in Reddy et al. (2012). The star's photometric temperature (T$^{\rm phot}_{\rm eff}$) was derived by substituting the dereddened\footnote{The adopted interstellar extinctions are (A$_{V}$, A$_{K}$, E(V-K), E(J-K))= (3.1, 0.28, 2.75, 0.54)*E(B-V)} photometric colours into the infrared flux method based colour$-$temperature calibrations of Alonso, Arribas \& Mart\'{i}nez-Roger (1999). Adopted mean values of interstellar reddening and metallicity are E(B-V)$=$0.12 (Johnson et al. 2008) and [Fe/H]$=-$1.4 dex (Calamida et al. 2017), respectively. Although Omega Cen has indisputable spread in metallicity (Norris \& Da Costa 1995; Johnson \& Pilachowski 2010; Villanova et al. 2014) and reddening (Calamida et al. 2005), consideration of their affect on T$^{\rm phot}_{\rm eff}$ is irrelevant given that our final chemical abundances will be  based on the spectroscopically measured effective temperature and surface gravity.

The surface gravity, $\log~g_{\rm phot}$, was determined by incorporating the distance modulus of (m-M)$_{\rm V}=$13.7 (van de Ven et al. 2006), photometric temperature, bolometric correction $BC_{V}$, red giants mass of $M=$\,0.80 M$_\odot$ (Johnson \& Pilachowski 2010) and T$_{\rm eff},_{\odot}$= 5777 K and log~$g_{\odot}$= 4.44 cm s$^{-2}$ of the Sun into the standard log~$g$\,$-$\,T$_{\rm eff}$ relation (Reddy et al. 2012). The values of $BC_{V}$ were obtained using equation 17 in Alonso et al. (1999).
The values of reddening and distance modulus adopted here are in fair agreement with the cluster mean reddening and distance modulus of 0.132 and 13.72, respectively, derived by Bono et al. (2019 and references therein) from the accurate and homogeneous optical (BVI) and infrared (JHK) mean magnitudes of RR Lyrae stars selected from the Gaia source catalog (Gaia Collaboration et al. 2016, 2018b).

A differential abundance analysis relative to the Sun was performed using the {\it abfind} driver of {\scs \bf MOOG}\footnote{{\scs \bf MOOG} was developed and updated by Chris Sneden and originally described in Sneden (1973)} adopting 1-dimensional theoretical model atmospheres and the iron line EWs following the local thermodynamic equilibrium (LTE) abundance analysis technique (Reddy et al. 2015). Photometric temperature and gravity was used to generate initial model photosphere via interpolating linearly within the ATLAS9 model atmosphere grid of Castelli \& Kurucz (2003). Reference solar abundances derived using solar EWs measured off the solar integrated disk spectrum (Kurucz et al. 1984) and the Kurucz model photosphere with T$_{\rm eff},_{\odot}$ = 5777 K, log~$g_{\odot}$ = 4.44 cm s$^{-2}$ and [Fe/H]$=$0.00 are taken from Reddy et al. (2012). 

Starting with initial theoretical model, the individual iron line abundances were force-fitted to match the computed EWs to observed ones by imposing three conditions: excitation and ionization equilibrium and the independent relation between iron line abundances and line's reduced EWs (Sneden 1973). The microturbulence velocity, $\xi_{t}$, assumed to be isotropic and independent of depth in the stellar atmosphere is derived by forcing no trend between the Fe abundance from Fe {\scs II} lines and the line's reduced EWs. The effective temperature is estimated by reducing the slope between the iron line abundances (log\,$\epsilon$(Fe {\scs I})) and line's lower excitation potential (LEP) to almost zero (excitation equilibrium).
The surface gravity, log~$g$, is adjusted in steps of $\pm$0.05 dex until the Fe abundance from Fe {\scs I} and Fe {\scs II} lines agree well within 0.02 dex (i.e., ionization equilibrium between the neutral and ionized species) for the derived $T_{\rm eff}$ and $\xi_{t}$. 
As the stellar parameters {\it T}$_{\rm eff}$, log~${g}$ and $\xi_{\rm t}$ are interdependent, several iterations are required to extract a suitable model from the grid of model photospheres.

The uncertainties in stellar parameters are derived following Reddy et al. (2015). The uncertainty in $T_{\rm eff}$ is estimated to be the temperature difference from the chosen model value that would introduce spurious trends in log\,$\epsilon$(Fe {\scs I}) versus line's LEP corresponding to the iron abundance deviating by 1$\sigma$ standard deviation of the derived average Fe abundance. This condition is also satisfied for lines of elements Ti, Cr and Ni covering a good range in line's LEP. The uncertainty in $\xi_{t}$ would be the difference in $\xi_{t}$ that would introduce spurious trends in log\,$\epsilon$(Fe {\scs II}) vs. reduced EWs and cause a variation in [Fe/H] value larger than 1$\sigma$. This condition is satisfied by other species such as Fe {\scs I}, Ni {\scs I}, and Cr {\scs I} covering a large range in line EWs (about 20 m\AA\ to 120 m\AA). The error in surface gravity is assumed to be the difference in log~$g$ from chosen stellar model that results about 1$\sigma$ difference between Fe {\scs I} and Fe {\scs II} line abundances. A satisfactory agreement of this aspect is provided by other elements whose abundance estimates are based on both neutral and ionized lines (e.g., Ti and Cr). The typical internal (star-to-star) errors derived in this process are 50 K in $T_{\rm eff}$, 0.1 dex each in log~$g$ and $\xi_{t}$. The stellar parameters estimated for program stars are given in Table \ref{stellar_param}.  

A comparison between photometric and spectroscopic atmospheric parameters is offered in table \ref{stellar_param}. With the exception of LEID 48319, spectroscopic $T_{\rm eff}$s and log~$g$s are in good agreement with photometric ones: Excluding two stars LEID 32007 and LEID 48319 having no 2MASS colours, mean difference in photometric temperatures estimated using (B-V) and (V-K) is $-$73 $\pm$ 189 K and using (V-K) and (J-K) is $+$88 $\pm$ 66 K. The corresponding mean differences between (B-V), (V-K) and (J-K) based colour temperatures and spectroscopic T$^{spec}_{\rm eff}$'s are $-$98 $\pm$ 145 K, $-$25 $\pm$ 68 K and $-$113 $\pm$ 45 K, respectively. No single colour-based temperature agrees well with spectroscopic temperatures for all stars. However, the temperature differences between photometric and spectroscopic estimates are within uncertainties found between different colour-based temperatures for the same star. The mean differences in gravities and luminosities across the sample of 11 stars (neglecting LEID 32007 and LEID 48319) are $+$0.07$\pm$0.19 dex and $-$0.12$\pm$0.14 dex, respectively.

The photometric temperature of one star (LEID 48319) differ by almost 870 K with respect to spectroscopic value derived using iron line EWs. LEID 48319 has no 2MASS magnitudes to cross check this result. Moreover, this star has no radial velocity estimates in the literature. Spectroscopic temperature derived in this paper is consistent with its (B-V) colour being at 1.335 i.e., about 0.44 away from its current value of 0.891 in Table \ref{log_observations}. Typical uncertainties in photometric measurements, as outlined below, are unable to account for such a large discrepancy in temperature. Nonetheless, the inclusion of this star in abundance analysis is not going to influence conclusions of this paper. 

The photometric stellar parameters estimated using the infrared flux method based colour$-$temperature relations are mainly sensitive to the adopted colours, reddening, metallicity and distance modulus of the cluster. A typical uncertainty of 0.02 mag. each in (B-V) and E(B-V), 0.05 dex in metallicity renders total errors of 65 K and 52 K in temperatures obtained from (B-V) and (V-K) relations, respectively. The (J-K) vs. T$_{\rm eff}$ relation is independent of metallicity, but errors of 0.02 mag. and 0.011 (i.e., 0.54*E(B-V)), respectively in colour and reddening adds a total temperature uncertainty of 68 K (Reddy \& Lambert 2019). Here the total error in T$_{\rm eff}$ is the quadratic sum of errors introduced by varying the parameters used in the photometric relations. Note, however, that these are the lower limits whose values tend to increase if true errors in photometry are adopted. Similarly, the above errors in T$_{\rm eff}$s and an uncertainty of 0.2 mag. in distance modulus yield an uncertainty of 0.08 dex in photometric surface gravities. The inclusion of differential reddening instead of adopting a single value for all stars in Table \ref{stellar_param} may further boost uncertainties in photometric stellar parameters.

\begin{table*}
\caption{Elemental abundance ratios [El/Fe] and iron abundance ([Fe I/H] and [Fe II/H]) for stars in Omega Cen analysed in this paper. Abundances calculated by synthesis are presented in bold typeface. Mg abundances will be presented in section \ref{helium_abundances}. }
\label{mean_abundance}
\begin{tabular}{lccccccc}   \hline
 \multicolumn{1}{c}{Species} & \multicolumn{1}{c}{26067} & \multicolumn{1}{c}{32149} & \multicolumn{1}{c}{33164} &
 \multicolumn{1}{c}{34029} & \multicolumn{1}{c}{34180} & \multicolumn{1}{c}{37024} & \multicolumn{1}{c}{41476}   \\ \hline

$[$O I/Fe$]$   &{\bf $+$0.29 }   &{\bf$+$0.74}     &{\bf$+$0.54 }    &{\bf$+$0.65 }    &{\bf$+$0.72 }    &{\bf$+$0.38 }    &{\bf$+$0.81 }   \\
$[$Na I/Fe$]LTE$  &$+$0.49$\pm$0.01 &$+$0.70$\pm$0.02 &$+$0.83$\pm$0.01 &$+$0.26$\pm$0.03 &$+$0.75$\pm$0.03 &$+$0.94$\pm$0.02 &$-$0.08$\pm$0.03 \\
$[$Na I/Fe$]NLTE$&$+$0.42$\pm$0.01 &$+$0.61$\pm$0.02 &$+$0.74$\pm$0.01 &$+$0.23$\pm$0.03 &$+$0.64$\pm$0.03 &$+$0.79$\pm$0.02 &$-$0.12$\pm$0.03 \\
$[$Al I/Fe$]LTE$  &{\bf$+$0.58$\pm$0.03} &{\bf$+$0.60$\pm$0.02} &{\bf$+$0.53$\pm$0.01} &{\bf$+$0.32$\pm$0.01} &{\bf$+$0.52$\pm$0.02} &{\bf$+$0.83$\pm$0.04} &{\bf$+$0.19$\pm$0.02} \\
$[$Al I/Fe$]NLTE$ &$+$0.47$\pm$0.03 &$+$0.49$\pm$0.02 &$+$0.40$\pm$0.01 &$+$0.19$\pm$0.01 &$+$0.40$\pm$0.02 &$+$0.69$\pm$0.04 &$+$0.08$\pm$0.02 \\
$[$Si I/Fe$]$  &$+$0.67$\pm$0.02 &$+$0.60$\pm$0.03 &$+$0.75$\pm$0.02 &$+$0.49$\pm$0.01 &$+$0.43$\pm$0.02 &$+$0.81$\pm$0.00 &$+$0.47$\pm$0.00 \\
$[$Ca I/Fe$]$  &$+$0.43$\pm$0.02 &$+$0.43$\pm$0.03 &$+$0.36$\pm$0.02 &$+$0.25$\pm$0.02 &$+$0.38$\pm$0.03 &$+$0.58$\pm$0.02 &$+$0.39$\pm$0.01 \\
$[$Ti I/Fe$]$  &$+$0.31$\pm$0.03 &$+$0.40$\pm$0.03 &$+$0.26$\pm$0.01 &$+$0.29$\pm$0.05 &$+$0.69$\pm$0.03 &$+$0.73$\pm$0.01 &$+$0.57$\pm$0.02 \\
$[$Ti II/Fe$]$ &$+$0.38$\pm$0.01 &$+$0.31$\pm$0.04 &$+$0.29$\pm$0.01 &$+$0.34$\pm$0.05 &$+$0.62$\pm$0.03 &$+$0.70$\pm$0.04 &$+$0.61$\pm$0.02 \\
$[$Cr I/Fe$]$  &$-$0.01$\pm$0.04 &$+$0.03$\pm$0.03 &$+$0.15$\pm$0.03 &$-$0.01$\pm$0.03 &$+$0.22$\pm$0.02 &$+$0.21$\pm$0.02 &$-$0.15$\pm$0.02 \\
$[$Cr II/Fe$]$ &$+$0.24$\pm$0.02 &$+$0.14$\pm$0.02 &$+$0.14$\pm$0.01 &$+$0.12$\pm$0.02 &$+$0.19$\pm$0.02 &      $\ldots$   &$-$0.13$\pm$0.02 \\
$[$Fe I/H$]$   &$-$0.80$\pm$0.03 &$-$0.74$\pm$0.04 &$-$0.89$\pm$0.03 &$-$1.31$\pm$0.03 &$-$0.75$\pm$0.03 &$-$0.93$\pm$0.03 &$-$1.17$\pm$0.03 \\
$[$Fe II/H$]$  &$-$0.79$\pm$0.02 &$-$0.74$\pm$0.02 &$-$0.91$\pm$0.02 &$-$1.29$\pm$0.01 &$-$0.75$\pm$0.03 &$-$0.92$\pm$0.02 &$-$1.17$\pm$0.03 \\
$[$Ni I/Fe$]$  &$-$0.01$\pm$0.02 & \,0.00$\pm$0.03 &$-$0.01$\pm$0.03 &$+$0.02$\pm$0.02 &$+$0.07$\pm$0.02 &$+$0.14$\pm$0.02 &$+$0.08$\pm$0.03 \\
$[$Y II/Fe$]$  &$+$0.77$\pm$0.02 &$+$0.82$\pm$0.02 &$+$0.54$\pm$0.02 &$+$0.69$\pm$0.01 &$+$0.76$\pm$0.03 &$+$0.86$\pm$0.02 &$+$0.99$\pm$0.02 \\
$[$Zr I/Fe$]$  &$+$0.68$\pm$0.01 &$+$0.98$\pm$0.02 &$+$0.61$\pm$0.05 &$+$0.45$\pm$0.02 &$+$0.90$\pm$0.03 &$+$1.72$\pm$0.02 &$+$0.94$\pm$0.01 \\
$[$Ba II/Fe$]$ &{\bf$+$0.84 }    &{\bf$+$0.92 }    &{\bf$+$0.80 }    &{\bf$+$0.79 }    &{\bf$+$0.95 }    &{\bf$+$1.02 }    &{\bf$+$0.77 }    \\
$[$La II/Fe$]$ &$+$0.70$\pm$0.03 &$+$0.85$\pm$0.02 &$+$0.75$\pm$0.03 &$+$0.66$\pm$0.03 &$+$0.72$\pm$0.00 &$+$0.95$\pm$0.03 &$+$0.49$\pm$0.00 \\
$[$Ce II/Fe$]$ &$+$0.70$\pm$0.03 &$+$0.68$\pm$0.03 &$+$0.73$\pm$0.02 &$+$0.62$\pm$0.03 &$+$0.77$\pm$0.02 &$+$0.89$\pm$0.00 &$+$0.58$\pm$0.01 \\
$[$Sm II/Fe$]$ &$+$0.70$\pm$0.02 &$+$0.84$\pm$0.02 &$+$0.79$\pm$0.01 &$+$0.67$\pm$0.02 &$+$0.78$\pm$0.02 &$+$0.92$\pm$0.02 &$+$0.79$\pm$0.02 \\
\hline
\multicolumn{8}{c}{ } \\
\multicolumn{1}{c}{Species} & \multicolumn{1}{c}{42042} & \multicolumn{1}{c}{48319} & \multicolumn{1}{c}{54022} &
\multicolumn{1}{c}{32007} & \multicolumn{1}{c}{49026} & \multicolumn{1}{c}{78035} & \multicolumn{1}{c}{ }  \\ \hline

$[$O I/Fe$]$   &{\bf$+$0.35 }    &{\bf$+$0.77 }    &{\bf$+$0.44 }    &{\bf$+$0.27 }    &{\bf$+$0.21 }    &{\bf$+$0.54 }    & \\
$[$Na I/Fe$]LTE$  &$+$0.60$\pm$0.03 &$+$0.82$\pm$0.03 &$+$0.81$\pm$0.01 &$+$0.19$\pm$0.04 &$+$0.32$\pm$0.02 &$-$0.04$\pm$0.03 & \\
$[$Na I/Fe$]NLTE$&$+$0.52$\pm$0.03 &$+$0.72$\pm$0.03 &$+$0.71$\pm$0.01 &$+$0.14$\pm$0.04 &$+$0.25$\pm$0.02 &$-$0.08$\pm$0.03 & \\
$[$Al I/Fe$]LTE$  &{\bf$+$0.68$\pm$0.04} &{\bf$+$0.52$\pm$0.01} &{\bf$+$0.70$\pm$0.02} &{\bf$+$0.36$\pm$0.02} &{\bf$+$0.41$\pm$0.02} &{\bf$+$0.31$\pm$0.01} & \\
$[$Al I/Fe$]NLTE$ &$+$0.56$\pm$0.04 &$+$0.40$\pm$0.01 &$+$0.58$\pm$0.02 &$+$0.26$\pm$0.02 &$+$0.31$\pm$0.02 &$+$0.20$\pm$0.01 & \\
$[$Si I/Fe$]$  &$+$0.57$\pm$0.02 &$+$0.49$\pm$0.02 &$+$0.58$\pm$0.02 &$+$0.35$\pm$0.02 &$+$0.40$\pm$0.02 &$+$0.45$\pm$0.02 & \\
$[$Ca I/Fe$]$  &$+$0.47$\pm$0.01 &$+$0.29$\pm$0.01 &$+$0.45$\pm$0.03 &$+$0.21$\pm$0.01 &$+$0.30$\pm$0.01 &$+$0.03$\pm$0.03 & \\
$[$Ti I/Fe$]$  &$+$0.22$\pm$0.01 &$+$0.60$\pm$0.02 &$+$0.48$\pm$0.01 &$+$0.28$\pm$0.01 &$+$0.25$\pm$0.02 &$+$0.15$\pm$0.01 & \\
$[$Ti II/Fe$]$ &$+$0.28$\pm$0.02 &$+$0.52$\pm$0.00 &$+$0.46$\pm$0.02 &$+$0.28$\pm$0.02 &$+$0.25$\pm$0.02 &$+$0.21$\pm$0.01 & \\
$[$Cr I/Fe$]$  &$-$0.01$\pm$0.02 &$+$0.39$\pm$0.00 &$+$0.28$\pm$0.02 &$+$0.16$\pm$0.02 &$+$0.19$\pm$0.02 &$-$0.03$\pm$0.03 & \\
$[$Cr II/Fe$]$ &$-$0.01$\pm$0.03 &      $\ldots$   &$+$0.30$\pm$0.01 &$+$0.17$\pm$0.02 &$+$0.16$\pm$0.02 &$-$0.04$\pm$0.01 & \\
$[$Fe I/H$]$   &$-$0.84$\pm$0.03 &$-$0.76$\pm$0.03 &$-$0.78$\pm$0.03 &$-$0.42$\pm$0.04 &$-$0.46$\pm$0.03 &$-$0.69$\pm$0.03 & \\
$[$Fe II/H$]$  &$-$0.83$\pm$0.02 &$-$0.75$\pm$0.02 &$-$0.77$\pm$0.02 &$-$0.41$\pm$0.03 &$-$0.45$\pm$0.03 &$-$0.67$\pm$0.02 & \\
$[$Ni I/Fe$]$  &$+$0.16$\pm$0.02 &$-$0.04$\pm$0.03 &$+$0.07$\pm$0.03 &$+$0.20$\pm$0.02 &$+$0.23$\pm$0.02 &$+$0.14$\pm$0.01 & \\
$[$Y II/Fe$]$  &$+$0.80$\pm$0.02 &$+$0.62$\pm$0.01 &$+$0.94$\pm$0.02 &$+$0.33$\pm$0.03 &$+$0.34$\pm$0.02 &$+$0.47$\pm$0.02 & \\
$[$Zr II/Fe$]$ &$+$0.69$\pm$0.02 &$+$1.01$\pm$0.02 &$+$1.12$\pm$0.00 &$-$0.07$\pm$0.02 &$-$0.04$\pm$0.02 &$+$0.08$\pm$0.01 & \\
$[$Ba II/Fe$]$ &{\bf$+$0.94  }   &{\bf$+$1.05  }   &{\bf$+$1.22 }    &{\bf$-$0.09 }    &{\bf$-$0.10 }    &{\bf$+$0.13 }    & \\
$[$La II/Fe$]$ &$+$0.75$\pm$0.01 &$+$0.84$\pm$0.03 &$+$1.10$\pm$0.02 &$-$0.19$\pm$0.00 &$-$0.10$\pm$0.04 &$+$0.06$\pm$0.02 & \\
$[$Nd II/Fe$]$ &$+$0.78$\pm$0.02 &$+$0.85$\pm$0.02 &$+$1.07$\pm$0.03 &$-$0.11$\pm$0.00 &\,0.00$\pm$0.02  &$-$0.01$\pm$0.02 & \\
$[$Sm II/Fe$]$ &$+$0.72$\pm$0.02 &$+$0.98$\pm$0.02 &$+$1.07$\pm$0.02 &$+$0.06$\pm$0.01 &$+$0.07$\pm$0.01 &$+$0.22$\pm$0.03 & \\

\hline
\end{tabular}
\end{table*}

The chemical abundances for other elements were derived using either line EWs or spectrum synthesis analyses adopting the spectroscopic stellar parameters derived previously using iron lines. The magnesium abundance was derived using both the atomic Mg and MgH line transitions with well calibrated linelists discussed below which give consistent [Mg/H] abundances from Mg {\scs I} and MgH lines for stars having the He/H$-$ratio consistent with model photosphere value. The abundances for elements affected by hyperfine structure (hfs) and isotopic shifts (e.g., Ba) are also based on the synthetic spectrum analysis. The suite of lines included in the synthetic spectrum analysis are O (6300 \AA), Mg, MgH, Al (5557, 6696, 6698 \AA), and Ba (5853 \AA). Throughout this paper, the standard LTE synthetic profile fitting technique was followed by running the {\it synth} driver of {\scs \bf MOOG}.

\begin{table*}
{\fontsize{8}{8}\selectfont
\caption{Sensitivity of derived abundances to uncertainties in stellar parameters for a representative star LEID 42042 with $T_{\rm eff}$= 4200 K, $\log{g}$= 1.4 cm s$^{-2}$,and $\xi_{t}$= 1.6 km s$^{-1}$. Each number refer to the difference between the abundances obtained with and without varying each stellar parameter separately, while keeping the other parameters unchanged. }
\label{sensitivity}
\begin{tabular}{cccccc}   \hline
\multicolumn{1}{c}{Species} &\multicolumn{1}{c}{T$_{\rm eff}\pm$50} &\multicolumn{1}{c}{$\log\,g\pm$0.1} & \multicolumn{1}{c}{$\xi_{t}\pm$0.1} &
\multicolumn{1}{c}{[M/H]$\pm$0.1} & \multicolumn{1}{l}{ } \\ \cline{2-5} 
\multicolumn{1}{c}{(X)}&\multicolumn{1}{c}{$\delta_{X}$} &\multicolumn{1}{c}{$\delta_{X}$} & \multicolumn{1}{c}{$\delta_{X}$} &
\multicolumn{1}{c}{$\delta_{X}$}  &\multicolumn{1}{c}{$\sigma_{2}$} \\ \hline

$[$O I/Fe$]$   & $-0.04/-0.02$  & $+0.02/-0.02$  & $+0.03/-0.05$  & $+0.01/-0.03$ & 0.06  \\
$[$Na I/Fe$]$  & $-0.01/-0.04$  & $-0.02/+0.02$  & $+0.00/-0.01$  & $-0.03/+0.02$ & 0.04  \\
$[$Al I/Fe$]$  & $-0.01/-0.03$  & $-0.01/+0.02$  & $+0.02/-0.03$  & $-0.03/+0.02$ & 0.04  \\
$[$Si I/Fe$]$  & $-0.09/+0.05$  & $+0.00/+0.00$  & $+0.05/-0.03$  & $+0.00/+0.01$ & 0.08  \\
$[$Ca I/Fe$]$  & $+0.00/-0.05$  & $-0.01/+0.02$  & $-0.03/+0.00$  & $-0.02/+0.03$ & 0.04  \\
$[$Ti I/Fe$]$  & $+0.01/-0.06$  & $+0.00/+0.01$  & $-0.01/-0.01$  & $-0.02/+0.01$ & 0.04  \\
$[$Ti II/Fe$]$ & $-0.06/+0.02$  & $+0.02/-0.01$  & $+0.01/-0.02$  & $+0.01/+0.03$ & 0.05  \\
$[$Cr I/Fe$]$  & $-0.01/-0.04$  & $+0.00/+0.02$  & $+0.01/-0.03$  & $-0.02/+0.01$ & 0.04  \\
$[$Cr II/Fe$]$ & $-0.10/+0.06$  & $+0.02/-0.01$  & $+0.05/-0.04$  & $+0.01/+0.00$ & 0.09  \\
$[$Fe I/H$]$   & $+0.05/+0.00$  & $+0.01/-0.02$  & $-0.04/+0.06$  & $+0.03/-0.01$ & 0.06  \\
$[$Fe II/H$]$  & $-0.07/+0.08$  & $+0.03/-0.04$  & $+0.00/+0.05$  & $+0.06/-0.01$ & 0.09  \\
$[$Ni I/Fe$]$  & $-0.06/+0.00$  & $+0.00/-0.01$  & $-0.01/-0.01$  & $+0.00/-0.01$ & 0.03  \\
$[$Y II/Fe$]$  & $-0.05/+0.01$  & $+0.03/-0.01$  & $+0.00/+0.00$  & $+0.01/-0.01$ & 0.04  \\
$[$Zr I/Fe$]$  & $+0.03/-0.09$  & $+0.01/+0.00$  & $-0.03/-0.01$  & $-0.02/+0.00$ & 0.06  \\
$[$Ba II/Fe$]$ & $-0.05/-0.01$  & $+0.01/-0.01$  & $-0.07/+0.03$  & $+0.01/-0.03$ & 0.06  \\
$[$La II/Fe$]$ & $-0.04/-0.01$  & $+0.02/-0.02$  & $-0.01/-0.02$  & $+0.01/-0.02$ & 0.04  \\
$[$Ce II/Fe$]$ & $-0.05/+0.00$  & $+0.02/-0.02$  & $+0.01/-0.04$  & $+0.01/-0.03$ & 0.05  \\
$[$Nd II/Fe$]$ & $-0.04/-0.01$  & $+0.03/-0.02$  & $+0.01/-0.03$  & $+0.01/-0.02$ & 0.04  \\

\hline
\end{tabular}
} \vspace{-0.4cm}
\end{table*} 

The chemical abundances for individual red giants averaged over all available lines of given species are presented in Table \ref{mean_abundance} relative to solar abundances derived from the adopted $gf$-values (see Table 4 from Reddy et al. 2012). Entries in the table provide the average [Fe/H] and [X/Fe] for all elements (X) and standard deviation ($\sigma_{1}$). Following Reddy et al. (2015), the sensitivity of derived abundance ratios [El/Fe] to uncertainties in stellar parameters were derived by varying one of the stellar parameters of a model, keeping others fixed, by an amount equal to typical error mentioned earlier. The abundance differences ($\langle$[X/Fe]$\rangle$$_{new}$-$\langle$$[X/Fe]$$\rangle$) caused by variation of stellar parameters from the best model values are summed in quadrature to obtain a global uncertainty $\sigma_{2}$ (Table \ref{sensitivity}). The total internal error $\sigma_{tot}$ in [X/Fe] for each of the element is the quadratic sum of $\sigma_{1}$ and $\sigma_{2}$.

\begin{figure*}
\begin{center}
\includegraphics[trim=0.01cm 7.3cm 4.5cm 4.5cm, clip=true,height=0.30\textheight,width=0.70\textheight]{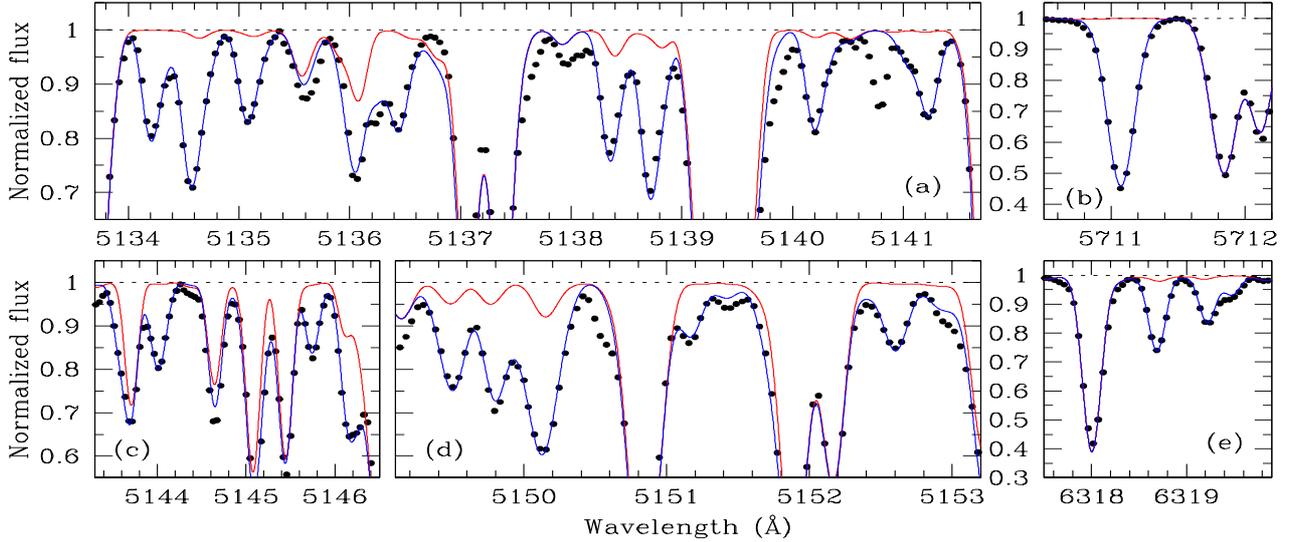}
\caption[]{Comparison of synthetic spectra (blue and red lines) with the observed spectra (black filled circles) of HD 17925 near the MgH lines in panels (a), (c), (d) and the Mg {\scs I} lines in panels (b) and (e). The best fit (blue line) to MgH and Mg lines and contaminating features from other elements (red line) are shown in each panel. }
\label{calib_HD17925}
\end{center}
\end{figure*}

\begin{figure*}
\begin{center}
\includegraphics[trim=0.05cm 10.4cm 0.2cm 4.3cm, clip=true,height=0.24\textheight,width=0.84\textheight]{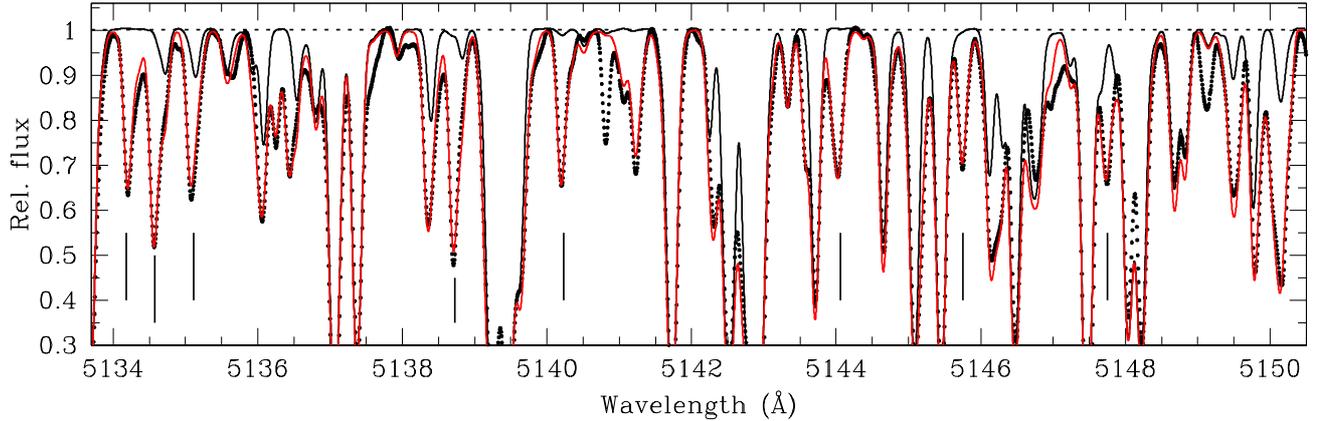} \vspace{-0.6cm}
\caption[]{Comparison of observed and synthetic spectra (solid black and red lines) for prominent MgH lines in the spectrum of the giant star Arcturus (black filled circles). The best fit synthetic spectrum (solid red line) to MgH lines and the contribution from lines of other elements are shown (solid black line). The central wavelengths of some of the MgH lines are marked with vertical lines. }
\label{synth_arcmgh}
\end{center}
\end{figure*}

\begin{figure}
\begin{center}
\includegraphics[trim=0.05cm 7.5cm 9.1cm 3.9cm, clip=true,height=0.27\textheight,width=0.46\textheight]{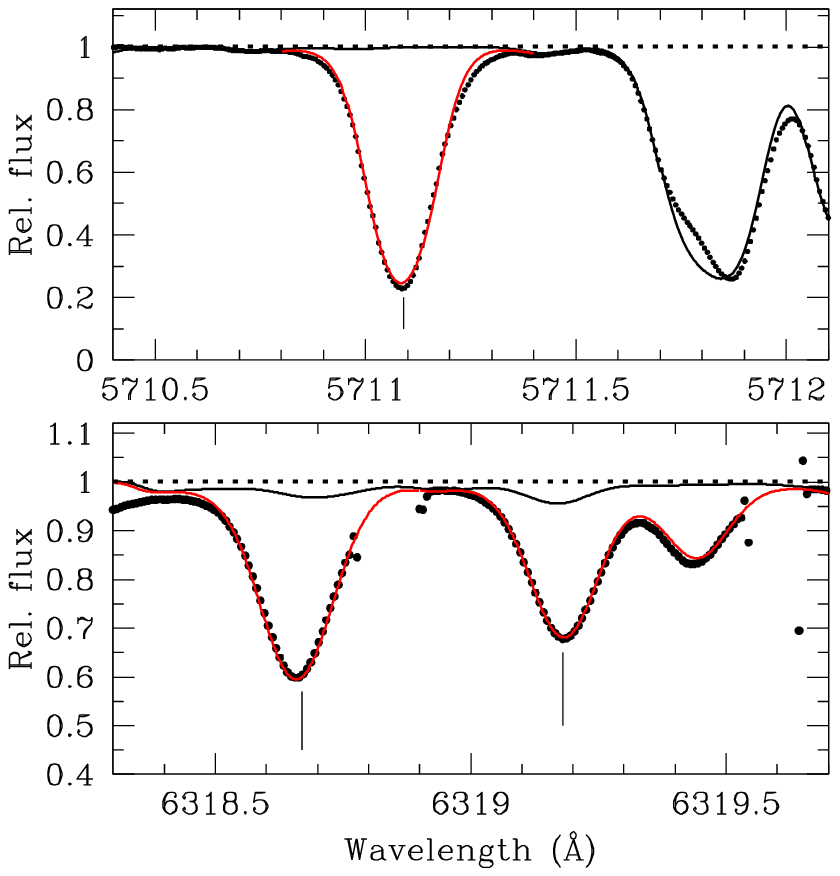} \vspace{-0.6cm}
\caption[]{Same as the Figure \ref{synth_arcmgh} but for the synthesis of atomic Mg lines at 5711, 6318.8 and 6319.2 \AA. }
\label{synth_arcmg}
\end{center}
\end{figure}

\subsection{Validation of linelists for MgH and Mg}  \label{validation_mg_mgh}
As the goal of this paper is to provide the best measure of He content of selected red giants in Omega Cen via the synthesis of lines of atomic Mg and molecular MgH, it is required to have a well calibrated linelists for spectrum synthesis analysis. 
The traditional method of validating atomic and molecular line data uses reproducing either the solar or Arcturus spectrum at selected key wavelengths for values of [X/H] for an element X that matches closely with published chemical composition of the Sun (Asplund et al. 2009) or Arcturus (Fulbright, McWilliam \& Rich 2007). 
The calibration of linelist against the solar spectrum is avoided for the reason that the theoretical EWs of the strongest MgH lines found between 5130 $-$ 5160 \AA\ region (e.g., 5134.2, 5134.6, 5136.0, 5138.8, 5140.2, 5141.2, 5144.1 \AA) are smaller than 3$-$5 m\AA\ and sensitive to continuum placement to derive a meaningful Mg abundance from the synthesis of MgH lines.

Validation of linelist against the spectrum of Arcturus (a more evolved giant) is discarded as well due to helium enrichment of its stellar atmosphere via the dredge-up episodes and helium flash during RGB phase that may results average [Mg/H] abundance from MgH lines smaller than that derived from the atomic Mg lines, if the adopted theoretical photosphere model of Arcturus has a solar helium-to-hydrogen ratio. This, as noted in the introduction, is due to decrease in the electron-to-gas pressure ratio when the He content is increased. The net effect of this is the reduction by same amount the dominant opacity source (i.e., H$^{-}$) in the stellar atmosphere, leading to line strength variations ( Maeckle et al. 1975; Piotto et al. 2005). Therefore, a wrong value of helium abundance adopted in models falsify the Mg abundance derived from MgH and Mg lines (Maeckle et al. 1975). This aspect will be investigated for Arcturus after validating the spectrum synthesis linelists in key MgH and Mg regions against calibration stars free of the issues mentioned earlier.
 
In this regard, the cool main-sequence stars observed at reasonably high spectral resolution and S/N-ratios provide spectra favourable for validating the linelists useful for the spectrum synthesis of targets in Omega Cen. Unlike the red giants, stellar atmospheres of dwarf stars are unaffected by mixing mechanisms which are expected to alter composition of He and other metals (C, N) including their isotopes in stellar atmosphere during the RGB phase (Charbonnel, Brown \& Wallerstein 1998). The spectra of a solar metallicity dwarf star -- HD 17925 -- observed at a resolution of 55,000 and S/N$>$ 200 with the Robert G. Tull coud\'{e} cross-dispersed echelle spectrograph (Tull et al. 1995) at the 2.7-m Harlan J. Smith reflector of the McDonald observatory is taken from Reddy \& Lambert (2015) for the purpose of validating Mg and MgH linelists. This star was analysed comprehensively for many elements from Na to Eu with typical measurement errors of less than 0.05 dex in [X/H]. 

Synthetic profiles were computed adopting the Mg isotopic ratios (Lodders 2003) and molecular line data from Hinkle et al. (2013) for MgH while the atomic line data for Mg lines was extracted from Ramirez \& Allende Prieto (2011) whose line selection comes from Asplund et al. (2009). The values of [Mg/H] derived in this paper from the synthesis of Mg lines at 5711, 6318.8, and 6319.2 \AA\ agree well with values reported by Reddy \& Lambert (2015). As shown in the Figure \ref{calib_HD17925}, a good fit to the MgH features (panels (a), (c), (d)) with values of [Mg/H] consistent with those derived from the atomic Mg lines (panels (b), (e)) validates the linelists adopted for the synthesis of Mg and MgH lines. From the Figure \ref{calib_HD17925}, the prominent MgH lines (5134.2, 5134.6, 5135.1, 5136.5, 5138.8, 5140.2, 5141.2, 5144.1, 5145.7, 5149.5, 5149.8, 5150.1, 5151.2, and 5152.6 \AA) of interest useful for Mg abundance estimates of stars are almost free of line contamination from other elements.

A further validation of Mg and MgH linelists is provided from the spectrum synthesis of Arcturus whose atmosphere suffers from helium enrichment due to its current evolutionary phase as an evolved red giant (Maeckle et al. 1975). Synthetic profiles of Mg and MgH lines computed with stellar parameters T$_{\rm eff}=$\,4286$\pm$ 30 K, log~${g}=$\,1.66 dex, $\xi_{t}=$\,1.74 km s$^{-1}$ and [Fe/H]$=-$\,0.52 dex (Ramirez \& Allende Prieto 2011) for the isotopic ratios of $^{24}$Mg:$^{25}$Mg:$^{26}$Mg=80:10:10 (Hinkle et al. 2013) and solar He/H fraction of 0.085 (Castelli \& Kurucz 2003) are compared with the NOAO Arcturus Spectral Atlas (Hinkle et al. 2000). Figure \ref{synth_arcmgh} and Figure \ref{synth_arcmg} shows the observed and computed spectra for prominent MgH and Mg lines in Arcturus. For the three atomic Mg lines in common, the derived average value of [Mg/H]$=$-0.11 dex is in fair agreement with a value of $-$0.13 dex in Ramirez \& Allende Prieto (2011). The best fit average value of [Mg/H]$=-$0.25 dex derived from the MgH lines is less than that obtained for the atomic Mg lines in Arcturus\footnote{Note, however, that Maeckle et al. (1975) noted a satisfactory agreement of Mg abundance calculated from the Mg {\scs I} and MgH lines in Arcturus for values of T$_{\rm eff}=$ 4260 K and log~${g}=$ 0.90 dex. Had I adopt their values of stellar parameters with the recent linelists discussed earlier, the fair agreement found by Maeckle et al. would be disturbed to give [Mg/H] from MgH lines larger by about 0.2 dex than the atomic Mg lines. Such a large positive abundance difference between MgH and Mg {\scs I} is caused mainly due to the greater sensitivity of molecular lines to surface gravity with the theoretical line strength decreases with gravity such that Mg abundance should be raised to match the MgH lines in Arcturus. Therefore, the lower value of [Mg/H] obtained for the lines of MgH relative to Mg {\scs I} in this paper is not an error of the abundance analysis.}.

The abundance difference for [Mg/H] between the Mg and MgH line abundances is larger than the total measurement error of about 0.07 dex (quadratic sum of errors on Mg and MgH line abundances) expected due to uncertainties in stellar parameters of Arcturus and the line-to-line scatter. Increasing slightly the theoretical Arcturus model value of He/H fraction to 0.16$\pm$0.02 renders satisfactory agreement of magnesium abundance derived from the Mg {\scs I} and MgH molecular lines i.e., similar value of [Mg/H] results from the atomic and molecular lines of Mg. This value of He/H$-$ratio is 0.06 larger, with 3$\sigma$ confidence level, than the solar He and H mass fractions adopted in Maeckle et al. (1975). If the currently accepted solar (primordial) He/H$-$ratio of 0.085 ($Y=$\,0.25, Castelli \& Kurucz 2003; Izotov, Thuan \& Stasi\'{n}ska 2007) is adopted, the level of confidence of the derived He/H$-$ratio in the atmosphere of Arcturus is more than 3$\sigma$. The theoretical stellar evolution models predict for a star of near solar-metallicity an enhancement of surface He/H$-$ratio by 0.015 after the first dredge-up whose value progressively increases along the RGB and during helium flash (Sweigart \& Gross 1978; Lattanzio 1986; Sweigart 1997).
This result, as anticipated, for Arcturus further validates the linelists adopted in the analysis of Mg and MgH lines and gives confidence that red giants with He-enhanced photospheres can be easily identified through the synthesis of atomic and molecular features of Mg in the observed spectra.

Note, however, that all the 1D model atmospheres (e.g., ATLAS, MARCS) based on the mixing-length theory (B\"{o}hm-Vitense 1958) of convective energy transport fail to represent the dynamic and multi-dimensional nature of convection in stellar atmospheres.
The full 3D model atmospheres compared to 1D are expected to represent both the velocity fields and thermal structure of the photosphere realistically and can have a dramatic impact on the formation of atomic and molecular lines in dwarfs and giants (Collet, Asplund \& Trampedach 2007; Asplund \& Garc\'{i}a P\'{e}rez 2001; Gallagher et al. 2016a; Thygesen et al. 2017).

In the 3D case for dwarfs, the region of line formation extends over a larger optical depth and peaks in shallower layers than when using traditional 1D atmospheres, with the contribution functions becoming shallower and broader with decreasing metallicity. The difference in temperature structure between 3D and 1D models are more substantial in the outer atmospheric layers of only low metallicity dwarfs ([Fe/H]$>$\,$-$2.0 dex, Collet et al. 2007; Gallagher et al. 2016a; Thygesen et al. 2017). These outer layers form molecules efficiently as they are generally cooler in 3D, resulting in stronger features that become more pronounced as the metallicity decreases. At metallicities higher than $-$1.0 dex, the line contribution functions in 3D models closely resemble the 1D case. The Mg (and C) abundances derived for solar metallicity dwarfs using the lines of molecular MgH (and CH) (Gallagher et al. 2016a; Thygesen et al. 2017) and atomic Mg (and C) (Collet et al. 2007; Bergemann et al. 2017) are very similar between the 3D and 1D model atmosphere analyses. Typically large abundance differences between 3D and 1D models are suggested for dwarf stars of metallicity less than $-$1.0 dex. Therefore, the 1D LTE abundances derived here for the solar-metallicity dwarf HD 17925 are unchanged when 3D model atmospheres are adopted.

In the case of giant stars with [Fe/H]$>-$\,1.0 dex, the differences in temperature structure and depth of formation of MgH lines are negligible between the 3D and 1D model atmospheres. Thygesen et al. (2017) showed that as the metallicity is increased in giant model atmospheres MgH lines begin to form over the same regions in the 3D atmosphere as they do in the equivalent 1D atmosphere, whereas in the metal-poor atmosphere ([Fe/H]$<$\,$-$1.0 dex) the MgH lines form further out in 3D. For the giants, the line equivalent width contribution functions are also notably shallower at high metallicity and the MgH lines in 3D models are slightly weaker than their 1D counterpart. Thygesen et al. suggested a positive correction ($>+$0.1 dex) for the LTE Mg abundances derived from the MgH lines adopting 1D model atmospheres for giants of metallicity [Fe/H]$>-$\,1.0 dex. The atomic Mg line formation in 3D is explored in Collet et al. (2007), who suggested negligible differences in atomic Mg line abundances between the 3D and 1D LTE atmospheres for metallicities between 0.0 and $-$\,1.0 dex. These results for red giants suggest that using 3D model atmospheres in place of 1D atmospheric models alleviate the difference in [Mg/H] abundances between Mg and MgH lines observed for Arcturus. 
As large libraries of 3D model atmospheres for stellar spectral analyses are not yet available, this paper explores the line abundances of elements including the atomic Mg and molecular MgH in red giants adopting 1D LTE Kurucz model photospheres.

\section{Non-LTE corrections} \label{nonlte_na_mg_al_fe}
In a recent study, Dupree et al. (2011) noted that He-enhanced giants are associated with enhancements of LTE [Na/Fe] and [Al/Fe] abundances. 
This section describes whether non-LTE effects may alter LTE abundances of Fe, Na and Al including the LTE Mg abundance in metal-poor giants of this paper.

The iron abundance derived assuming LTE was corrected for non-LTE effects using the grids of Lind et al. (2011). The non-LTE corrections were computed for a representative set of iron lines (Fe{\scs I}: the 6065.5, 6173.3, 6252.6, 6498.9, 6574.2, 6609.1, 6739.5, 6750.1; Fe\,{\sc ii}: 5197.6, 5325.5, 5425.3, 6369.5 lines) for the stellar parameters of three giants (LEID 26067, 37024 and 54022) representative of red giants in this paper. For the Sun, the averaged non-LTE Fe\,{\sc i} and Fe\,{\sc ii} abundance corrections are $+$0.02 dex and 0.00 dex, respectively. The average iron abundance corrections derived for the cluster giants are $+$0.04 dex and $-$0.01 dex, respectively, using the Fe\,{\sc i} and Fe\,{\sc ii} lines. These small corrections to LTE Fe abundances listed in Table \ref{mean_abundance} increases $[$Fe I/H$]$ by $+$0.02 dex, decreases $[$Fe II/H$]$ by $-$0.01 dex and cause a small downward revision of LTE abundances ([X/Fe]) of other elements by $+$0.02 dex. 
These non-LTE iron abundance corrections do not affect the [Fe/H] and [X/Fe] abundances presented in Table \ref{mean_abundance}, as the related correction is small (though not zero).

Sodium abundances derived assuming LTE were corrected for non-LTE effects on a line-by-line basis using the stellar parameters (e.g., T$_{\rm eff}$, log~$g$) and EWs of three Na lines (5688.20, 6154.22 and 6160.74 \AA) of each star as input to the interactive non-LTE database\footnote{\url{http://www.inspect-stars.com/}}. The non-LTE corrections were derived by interpolating to atmospheric parameters of program stars from the grids of Lind et al. (2011).

The non-LTE sodium abundance averaged over three Na lines for all stars including the Sun is lower than average LTE Na abundance by values in the range $-$0.09 to $-$0.21 dex. The average non-LTE correction obtained for the Sun using the above set of three Na lines is $-$0.09 dex. As a result, the non-LTE Na abundances derived relative to the Sun are not very much different from the differential LTE Na abundances. The non-LTE corrections applied for giants in this study reduce the differential LTE Na abundance on an average by 0.0 to $-$0.12 dex.

The non-LTE corrections for Al were computed using the extensive grids of abundance corrections discussed in Nordlander \& Lind (2017). The non-LTE Al abundances were calculated using the stellar parameters and LTE Al abundances obtained for the 6696.0 and 6698.7 \AA\ lines as input to the source code\footnote{Available online at \url{https://www.mso.anu.edu.au/~thomasn/NLTE/}} of Nordlander \& Lind. The averaged non-LTE Al abundance corrections derived for the Sun and cluster giants are $-$0.01 dex and $-$0.09 dex to $-$0.13 dex, respectively. These corrections for the cluster giants lower average LTE Al abundances by values in the range $-$0.08 to $-$0.12 dex. The non-LTE [X/Fe] ratios for Na and Al corrected for non-LTE [Fe/H] correction of $+$0.02 dex are listed in Table \ref{mean_abundance}. 

At the metallicity of cluster giants, all the Mg lines (5711, 6318.8, and 6319.2 \AA) including relatively strong Mg {\scs I} 5711 \AA\ line are quite insensitive to non-LTE effects (Osorio et al. 2015). The smallest non-LTE correction (about $-$0.01 dex) found for the 5711 \AA\ line for three representative giants (LEID 26067, 37024 and 54022) suggests that the non-LTE [Mg/H] abundance averaged over three lines of Mg is not different from LTE value of [Mg/H]. Therefore, no non-LTE corrections are applied for the LTE [Mg/H] values listed in Table \ref{abu_helium}.

\section{Results}
\subsection{Comparisons with the literature} 
Some of the stars analysed in this paper have been subjected to chemical composition analyses in the literature for a few elements using the low to medium-resolution spectra. The literature studies include, as noted in the Section \ref{obs_radial_velocities}, Simpson \& Cottrell (2013), Johnson \& Pilachowski (2010) and Marino et al. (2011). 

For stars and elements in common, the values of Simpson \& Cottrell (2013) for T$_{\rm eff}$, log~$g$, [Fe/H], [O/Fe], and [Ba/Fe] differ from this study by a mean of $-$25$\pm$63 K, $+$0.06$\pm$0.06 cm s$^{-2}$, $-$0.18$\pm$0.19 dex, $-$0.88$\pm$0.25 dex, and $-$0.25$\pm$0.12 dex, respectively. The mean difference in T$_{\rm eff}$, log~$g$, [Fe/H], [O/Fe], [Na/Fe], [Si/Fe], [Ca/Fe], [Ti/Fe], [Ni/Fe], and [La/Fe] values between Johnson \& Pilachowski (2010) and this work are $+$22$\pm$34 K, $-$0.01$\pm$0.09 cm s$^{-2}$, $-$0.03$\pm$0.13 dex, $-$0.4$\pm$0.1 dex, $+$0.27$\pm$0.17 dex, $+$0.19$\pm$0.26 dex, $+$0.09$\pm$0.24 dex, $-$0.04$\pm$0.25 dex, $+$0.02$\pm$0.05 dex, $-$0.29$\pm$0.20 dex, respectively. Marino et al. (2011)'s values for T$_{\rm eff}$, log~$g$, [Fe/H], [O/Fe], [Na/Fe], [Ba/Fe], and [La/Fe] differ in mean by $+$24$\pm$27 K, $+$0.13$\pm$0.03 cm s$^{-2}$, $-$0.04$\pm$0.11 dex, $-$0.25$\pm$0.11 dex, $+$0.17$\pm$0.15, $-$0.33$\pm$0.13 dex, $-$0.15$\pm$0.10 dex, respectively. 

For one of the stars LEID 41476, in common between Johnson \& Pilachowski (2010) and Marino et al. (2011), the stellar parameters are in good agreement within measurement errors with differences of $\Delta$T$_{\rm eff}=+$42 K and $\Delta$log~$g=-$0.26 cm s$^{-2}$, while larger discrepancies are noticed for Fe, and Na with values of $\Delta$[Fe/H]$=-$0.38 dex and $\Delta$[Na/Fe]$=+$0.52 dex, respectively. For four stars in common with this study (LEID numbers 26067, 41476, 42042 and 54022), the mean difference of values between Simpson \& Cottrell (2013) and Marino et al. (2011) for T$_{\rm eff}$, log~$g$, [Fe/H], [O/Fe], and [Ba/Fe] are $-$14$\pm$23 K, $-$0.03$\pm$0.03 cm s$^{-1}$, $-$0.19$\pm$0.24 dex, $-$0.51$\pm$0.29 dex, and $+$0.08$\pm$0.04 dex, respectively. Similarly, for the only star LEID 41476 in common between Johnson \& Pilachowski (2010) and Simpson \& Cottrell (2013), the stellar parameters and abundances for [Fe/H] and [O/Fe] differ by $\Delta$T$_{\rm eff}=-$30 K, $\Delta$log~$g=+$0.20 cm s$^{-2}$, $-$0.09 dex and $-$0.16 dex, respectively.

\begin{table*}
{\fontsize{8}{8}\selectfont
\caption{The average of [Mg/H] ratios from Mg (fifth column) and MgH (sixth column) lines for red giants analysed in this paper. The difference in average [Mg/H] between atomic Mg and MgH lines (seventh column) for He-normal (log\,$\epsilon$(He)$=$ 10.930 dex, $Y=$ 0.254) and He-enhanced models is listed along with true values of [Mg/Fe] derived adopting the He content. The significant positive difference in [Mg/H] ratios for all but three giants (LEID 33164, 34029, and 41476) strongly indicate He-enhanced nature of stellar atmospheres. }
\label{abu_helium}
\begin{tabular}{cccccccccc}   \hline
\multicolumn{1}{c}{Star} $\lvert$ & \multicolumn{4}{c}{[Mg/H] from Mg lines} $\lvert$ & \multicolumn{1}{c}{MgH lines} & \multicolumn{3}{c}{ } \\ \cline{2-5} 

\multicolumn{1}{c}{ } $\lvert$ & \multicolumn{1}{c}{$\lambda$ 5711} &\multicolumn{1}{c}{$\lambda$ 6318.8} & \multicolumn{1}{c}{$\lambda$ 6319.2 } & \multicolumn{1}{c}{[Mg/H]$_{\mbox{Avg.}}$ } $\lvert$ & \multicolumn{1}{c}{[Mg/H]$_{\mbox{Avg.}}$ } & \multicolumn{1}{c}{diff. } & \multicolumn{1}{c}{[Mg/Fe]} & \multicolumn{1}{c}{He} & \multicolumn{1}{c}{ $Y$ }     \\ \hline

LEID 26067 & $-$0.07 & $-$0.05 & $-$0.08 & $-$0.07 & $-$0.35 & $+$0.28&        &        &       \\
           & $-$0.25 & $-$0.23 & $-$0.25 & $-$0.24 & $-$0.24 & \,0.00 &$+$0.56 & 11.275 & 0.430  \\
LEID 32149 & $+$0.08 & $+$0.07 & $+$0.09 & $+$0.08 & $-$0.14 & $+$0.22&        &        &       \\
           & $-$0.04 & $-$0.03 & $-$0.03 & $-$0.03 & $-$0.03 & \,0.00 &$+$0.71 & 11.188 & 0.381  \\
LEID 33164 & $-$0.39 & $-$0.37 &$\ldots$ & $-$0.38 & $-$0.39 & $+$0.01&$+$0.51 & 10.930 & 0.254  \\
LEID 34029 & $-$0.58 & $-$0.55 & $-$0.61 & $-$0.58 & $-$0.58 & \,0.00 &$+$0.72 & 10.930 & 0.254  \\
LEID 34180 & $-$0.06 & $-$0.08 & $-$0.07 & $-$0.07 & $-$0.27 & $+$0.20&        &        &       \\
           & $-$0.26 & $-$0.24 & $-$0.24 & $-$0.25 & $-$0.24 & $-$0.01&$+$0.51 & 11.149 & 0.360  \\
LEID 37024 & $-$0.08 & $-$0.10 & $-$0.09 & $-$0.09 & $-$0.33 & $+$0.24&        &        &       \\
           & $-$0.32 & $-$0.30 & $-$0.29 & $-$0.30 & $-$0.30 & \,0.00 &$+$0.62 & 11.242 & 0.411  \\
LEID 41476 & $-$0.57 & $-$0.58 & $-$0.57 & $-$0.57 & $-$0.56 & $-$0.01&$+$0.61 & 10.930 & 0.254  \\
LEID 42042 & $+$0.02 & $+$0.01 & $+$0.02 & $+$0.02 & $-$0.42 & $+$0.44&        &        &       \\
           & $-$0.22 & $-$0.18 & $-$0.22 & $-$0.21 & $-$0.22 & $+$0.01&$+$0.61 & 11.337 & 0.465  \\
LEID 48319 & $-$0.01 &  \,0.00 & $-$0.02 & $-$0.01 & $-$0.20 & $+$0.19&        &        &       \\
           & $-$0.16 & $-$0.14 & $-$0.15 & $-$0.15 & $-$0.15 & \,0.00 &$+$0.60 & 11.149 & 0.360  \\
LEID 54022 & $-$0.06 & $-$0.08 & $-$0.09 & $-$0.08 & $-$0.34 & $+$0.26&        &        &       \\
           & $-$0.26 & $-$0.18 & $-$0.26 & $-$0.23 & $-$0.23 & \,0.00 &$+$0.54 & 11.242 & 0.411  \\
LEID 32007 & $+$0.07 & $+$0.05 & $+$0.06 & $+$0.06 & $-$0.13 & $+$0.19&        &        &       \\
           & $-$0.05 & $-$0.06 & $-$0.04 & $-$0.05 & $-$0.05 & \,0.00 &$+$0.36 & 11.275 & 0.430  \\
LEID 49026 & $+$0.25 & $+$0.25 & $+$0.24 & $+$0.25 & $-$0.13 & $+$0.38&        &        &       \\
           & $+$0.01 & $+$0.01 & $+$0.01 & $+$0.01 & $+$0.01 & \,0.00 &$+$0.46 & 11.307 & 0.448  \\
LEID 78035 & $-$0.11 & $-$0.10 & $-$0.09 & $-$0.10 & $-$0.30 & $+$0.20&        &        &       \\
           & $-$0.27 & $-$0.24 & $-$0.24 & $-$0.25 & $-$0.25 & \,0.00 &$+$0.43 & 11.188 & 0.381  \\
\hline
\end{tabular}
} \vspace{-0.4cm}
\end{table*}

\begin{figure*}
\begin{center}
\includegraphics[trim=0.05cm 10.4cm 0.2cm 4.3cm, clip=true,height=0.24\textheight,width=0.84\textheight]{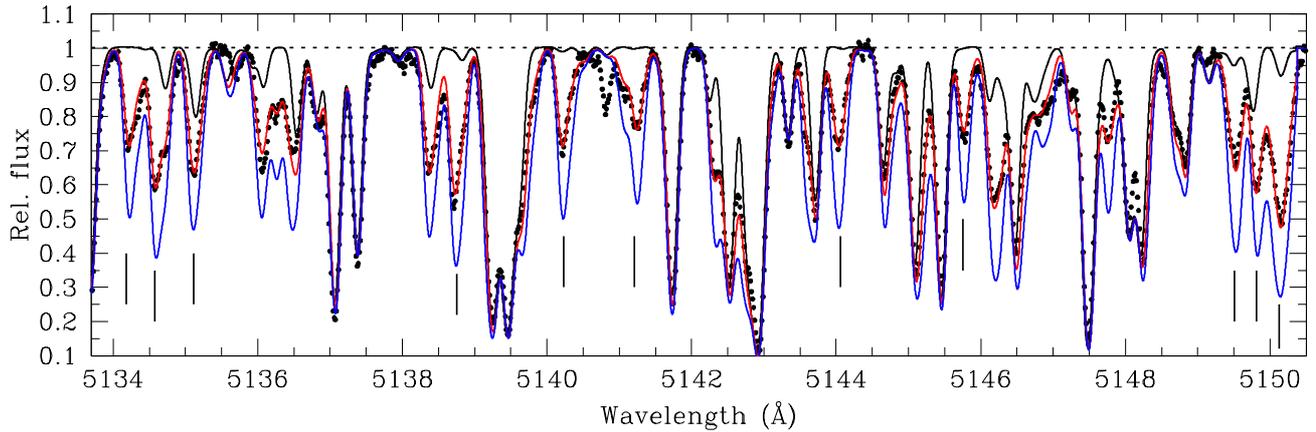} \vspace{-0.6cm}
\caption[]{Comparison of synthetic spectra (solid black, blue and red lines) with the observed spectrum (black filled circles) of LEID 42042 for the MgH lines. The solid red line corresponds to the best fit Mg abundance of [Mg/H]$=-$\,0.42 dex from the MgH lines. The expected strength of the MgH lines that corresponds to the Mg abundance derived using the atomic Mg lines in the observed spectrum (Figure \ref{synth_leid42042mg} and Table \ref{abu_helium}) is represented by a solid blue line. The solid black line is the synthetic spectrum generated, excluding the MgH lines in the linelist, for the spectral lines of other elements in the selected wavelength region.  The central wavelengths of some of the MgH lines are marked with vertical lines. }
\label{synth_leid42042mgh}
\end{center}
\end{figure*}

\begin{figure}
\begin{center}
\includegraphics[trim=0.05cm 7.5cm 9.1cm 3.9cm, clip=true,height=0.27\textheight,width=0.46\textheight]{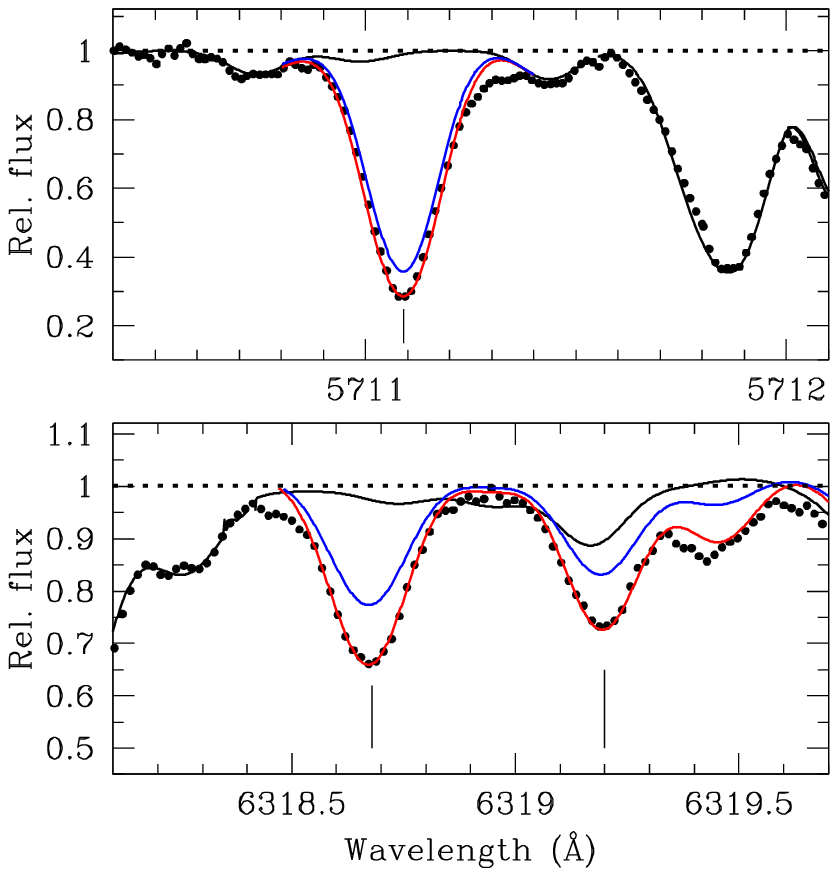} \vspace{-0.6cm}
\caption[]{ Same as figure \ref{synth_leid42042mgh} but for the synthesis of atomic Mg lines at 5711, 6318.8 and 6319.2 \AA\ in the observed spectrum of LEID 42042. The best fit synthetic spectrum to the atomic Mg lines is shown as a solid red line. As shown in the figure, the synthetic spectrum (solid blue line) computed for the Mg abundance derived from the MgH lines is inappropriate to match the atomic Mg line strengths and certainly underestimate the Mg abundance from atomic Mg lines. }
\label{synth_leid42042mg}
\end{center}
\end{figure}

The systematic offsets in stellar parameters among these studies are comparable to internal uncertainties in atmospheric parameters, while abundance estimates for some of the elements differ greatly between the analyses since different techniques have been used by different authors to derive stellar parameters and abundances (See, for example, Marino et al. 2011). Results found for giant stars in this paper are not considered exceptional given that within the similar metallicity range $-$0.5 to $-$1.3 dex, covered by the present sample, the difference in [X/Fe] abundances between Johnson \& Pilachowski (2010) and literature studies is of the order of 0.0$-$0.6 dex (see Figure 5 and 6 of Johnson \& Pilachowski 2010). This result signifies that systematic offsets in chemical abundances are typical of what is seen among similar abundance analyses by different authors of the same or similar stars. So, the abundance differences found here between this work and literature studies is not surprising. (Some of these differences are attributable to different values of solar abundances adopted in the literature studies.)

\subsection{Helium abundances} \label{helium_abundances}
Helium abundance of each red giant is estimated by minimizing the difference in [Mg/H] derived independently from the atomic Mg and MgH molecular lines in the observed spectra to less than 0.01 dex. Any deviation in value of helium abundance adopted in theoretical photospheric models from that observed in stellar atmosphere results inconsistent Mg abundances derived between atomic Mg and MgH lines. As noted in the introduction, an increase (decrease) in the He content of stellar atmosphere affects the continuum opacity due to hydrogen (H$^{-}$ ion) leading to weakening (strengthening) of MgH lines in the observed spectrum, while this effect is reversed for the atomic Mg lines i.e., weakening (strengthening) of MgH lines in He-enhanced atmosphere is associated with strengthening (weakening) of atomic Mg lines. Therefore, an incompatible (solar) helium abundance of the theoretical model results the average Mg abundance, as listed in Table \ref{abu_helium}, from the atomic Mg lines greater than the value measured from the synthesis of MgH lines in the observed spectra.

Synthetic profiles computed for the Mg isotopic ratios of $^{24}$Mg:$^{25}$Mg:$^{26}$Mg\,=\,70:15:15 (Da Costa, Norris \& Yong 2013) by varying the He/H$-$ratios of the adopted standard Kurucz model photosphere are matched simultaneously to the atomic Mg and MgH lines in the observed spectra. The adopted helium content of the star is the one that corresponds to the best fit value of He/H of the synthetic profile that minimizes the difference in average [Mg/H] values measured independently using the atomic Mg and MgH lines to less than 0.01 dex. An example synthetic profile fit to the MgH and Mg lines observed in the spectra of LEID 42042 is shown in Figure \ref{synth_leid42042mgh} and Figure \ref{synth_leid42042mg}. 

The inclusion of He-enhanced theoretical models computed for desired He/H$-$ratios instead of adopting the standard theoretical model photosphere (solar He/H$-$ratio) for the abundance analysis changes little the value of He/H derived using the standard stellar model from the synthesis of MgH and Mg lines in the observed spectra. This aspect is verified by calculating with ATLAS9 a new theoretical model photosphere with enhanced He content (i.e, log\,$\epsilon$(He)$=$\,11.337 dex) for the star LEID 42042, representative of red giants in this paper.
Synthetic profiles recomputed adopting this He-enhanced model are matched simultaneously to the atomic Mg and MgH lines in the observed spectra. A best fit to the lines of atomic Mg and molecular MgH for [Mg/H]$=-$\,0.22 dex requires a change in the He abundance of the model by $\Delta$log\,$\epsilon$(He)$=$\,0.06 dex ($\Delta$$Y=+$0.033).
These small corrections to the abundance and mass fraction of He derived from the observed spectra of red giants assuming He-normal photosphere models (Table \ref{stellar_param}) increases the values of log\,$\epsilon$(He) and $Y$ in the Table \ref{abu_helium} by 0.06 dex and 0.033, respectively. The LTE abundances of other elements are changed little ($+$0.03 dex or less). These small correction terms resulted adopting the He-enhanced theoretical models are in good agreement with previous analysis of stars in the globular cluster NGC 6121 by Villanova et al. (2012).

The internal error associated with the measured helium content was estimated by recomputing the synthetic profiles by varying the atmospheric parameters by their expected errors (as explained in the Section \ref{Stellar_parameters}) and matched to the observed spectra. The sensitivity of adopted helium abundance of a star to atmospheric parameters is deduced from the sensitivity of Mg and MgH lines to atmospheric parameters and the results are listed in Table \ref{sensitivity_He} for a representative star LEID 42042. The total uncertainty in the helium content obtained by summing in quadrature various contributions is $+$0.038 dex, dominated by mainly the temperature uncertainty. Error in microturbulence is negligible, while the surface gravity and metallicity only marginally contribute to the abundance uncertainty. The helium abundance of individual red giants is transformed to He mass fraction ($Y$) and the typical 1$\sigma$ error in $Y$ due to uncertainties in stellar parameters is $+$0.066. Systematic errors in atmospheric parameters with literature studies increases 1$\sigma$ error in $Y$ to 0.083. The helium abundance and mass fraction of helium for red giants in Omega Cen are presented in Table \ref{abu_helium}.

\begin{figure*}
\begin{center}
\includegraphics[trim=0.01cm 8.8cm 1.8cm 4.3cm, clip=true,height=0.30\textheight,width=0.70\textheight]{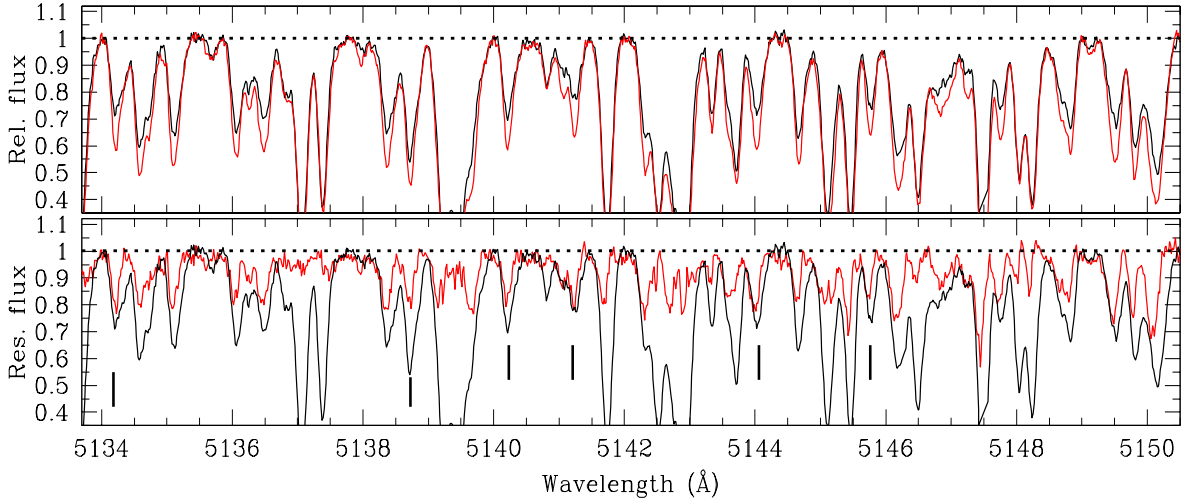} \vspace{-0.6cm}
\caption[]{Top panel: Comparison of relative strengths of MgH lines in the observed spectra of two red giants LEID 32149 (red, MgH strong) and LEID 42042 (black,  MgH weak).
Bottom panel: The residual flux shown as a red continuous line is resulted from the spectral division of LEID 32149 by LEID 42042. The observed spectrum of LEID 42042 (black) is overplotted for comparison. The central wavelengths of some of the MgH lines are marked with vertical lines. A novel result arriving from this plot and Figure \ref{comp_obs_spec_mg} is that the strength of MgH lines differ between stars having very similar set of atmospheric parameters (Table \ref{stellar_param}) and Mg line strengths, thus, acknowledge the reliability of relatively different Mg and He abundances derived using the MgH lines for giants in this paper (Table \ref{abu_helium}).}
\label{comp_obs_spec_mgh}
\end{center}
\end{figure*}

\begin{figure}
\begin{center}
\includegraphics[trim=0.01cm 8.8cm 6.8cm 4.3cm, clip=true,height=0.22\textheight,width=0.46\textheight]{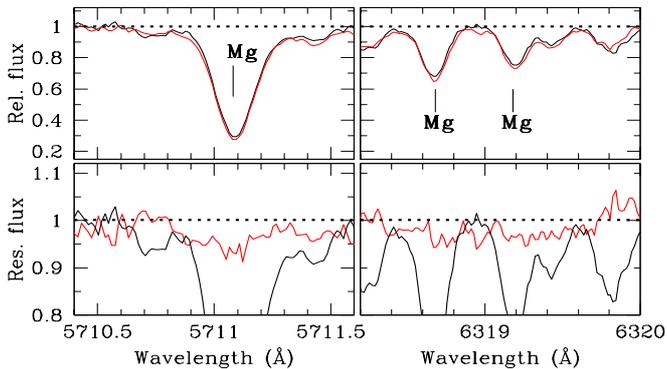} \vspace{-0.6cm}
\caption[]{Same as Figure \ref{comp_obs_spec_mgh} but for a comparison of relative strengths of Mg lines in the observed spectra of LEID 32149 (red) and LEID 42042 (black) }
\label{comp_obs_spec_mg}
\end{center}
\end{figure}

\begin{figure*}
\begin{center}
\includegraphics[trim=0.01cm 8.8cm 1.8cm 4.3cm, clip=true,height=0.30\textheight,width=0.70\textheight]{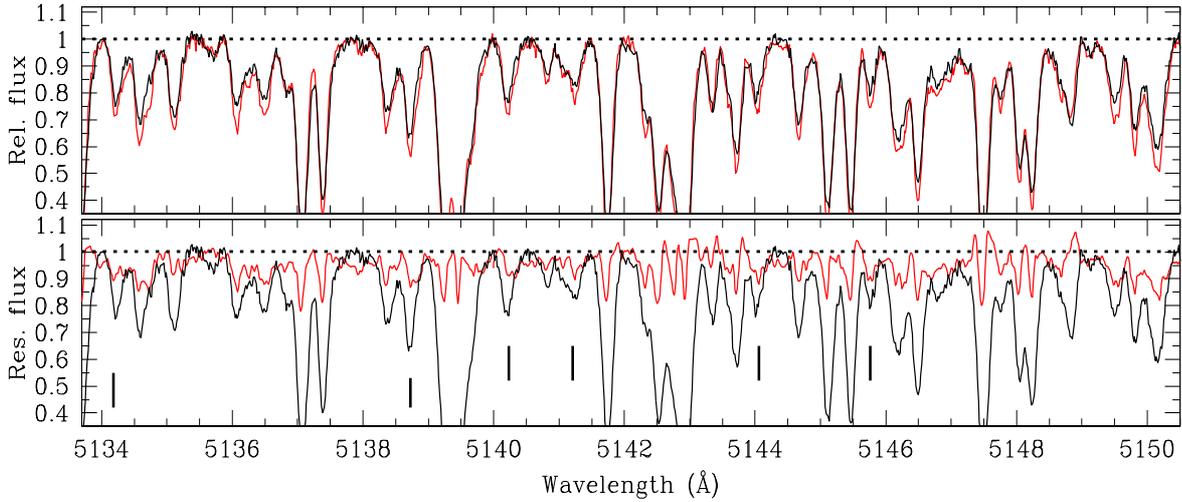} \vspace{-0.6cm}
\caption[]{Top panel: Comparison of relative strengths of MgH lines in the observed spectra of LEID 26067 (red, Mg {\scs I} strong) and LEID 33164 (black, Mg {\scs I} weak).
Bottom panel: The residual flux is resulted by the spectral division of LEID 26067 by LEID 33164. The observed spectrum of LEID 33164 (black) is overplotted for comparison. The central wavelengths of relatively clean MgH lines are marked with vertical lines. The difference in EWs for MgH lines is less than 5 m\AA, in spite of having very large differences in atomic Mg line strengths (Figure \ref{comp_l26067_l33164mg}).}
\label{comp_l26067_l33164mgh}
\end{center}
\end{figure*}

\begin{figure}
\begin{center}
\includegraphics[trim=0.01cm 8.8cm 6.8cm 4.3cm, clip=true,height=0.22\textheight,width=0.46\textheight]{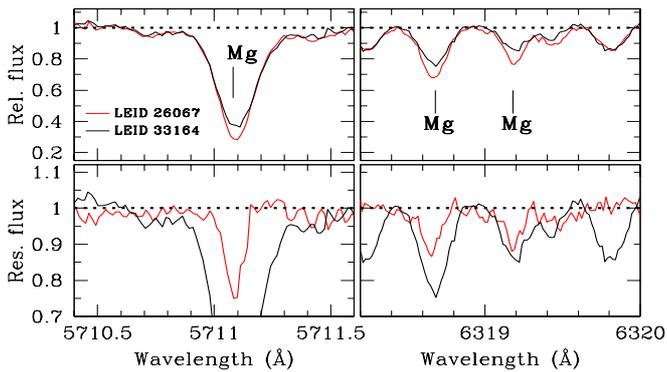} \vspace{-0.6cm}
\caption[]{Same as Figure \ref{comp_l26067_l33164mgh} but for a comparison of relative strengths of atomic Mg lines in observed spectra of LEID 26067 (red, Mg {\scs I} strong) and LEID 33164 (black, Mg {\scs I} weak). The Mg {\scs I} line in LEID 26067 is 20 m\AA\ stronger than that in LEID 33164. }
\label{comp_l26067_l33164mg}
\end{center}
\end{figure}

The reliability of the derived He content of the giants is probed through Figures \ref{comp_obs_spec_mgh}$-$\ref{comp_l26067_l33164mg} via a direct comparison of the relative strengths of atomic Mg and MgH lines in the observed spectra between two pairs of red giants with each pair represented by very similar set of stellar parameters (Table \ref{stellar_param}): LEID 32149 vs. LEID 42042 (Figure \ref{comp_obs_spec_mgh} and Figure \ref{comp_obs_spec_mg}) and LEID 26067 vs. LEID 33164 (Figure \ref{comp_l26067_l33164mgh} and Figure \ref{comp_l26067_l33164mg}). It is obvious from the Figure \ref{comp_obs_spec_mgh} that the strength of MgH lines (equivalent of Mg abundance from MgH lines) in LEID 32149 is about 1.2 times the strength of MgH lines in LEID 42042 (see the residual spectrum for comparison), whereas the line strengths of atomic Mg (equivalent of Mg abundance from Mg lines) differ little (Figure \ref{comp_obs_spec_mg}). The average EW of relatively clean MgH lines (marked with vertical lines in the Figure \ref{comp_obs_spec_mgh}) in LEID 32149 is about 25 m\AA\ greater than that in LEID 42042. From the Figure \ref{comp_l26067_l33164mgh} and Figure \ref{comp_l26067_l33164mg} shown for LEID 26067 and LEID 33164, MgH lines are very similar in strength, in spite of having significantly large differences in EWs of atomic Mg lines (i.e., the EWs of Mg lines in LEID 26067 are almost 20 m\AA\ larger than those in LEID 33164). Inspection of the Figures \ref{comp_obs_spec_mgh}$-$\ref{comp_l26067_l33164mg} confirms, between stars represented by similar set of stellar parameters, the existence of real differences in relative strengths (thus, the derived abundances) of either MgH lines for very similar Mg line EWs (seen for LEID 32149 and LEID 42042) or Mg lines for very similar MgH line EWs (LEID 26067 and LEID 33164).

In normal stellar atmospheres, hydrogen atoms always outnumber the available Mg atoms. In cool dwarfs and giants with conditions favourable for the formation of MgH molecules, it is expected for stars of similar stellar parameters that the MgH lines are very similar in strengths for similar Mg {\scs I} line profiles, and vice versa . Typically large inconsistency exceeding the measurement error in line strengths between stars for the MgH lines but not for the atomic Mg lines, and vice versa, signifies that the mass fraction of the source of opacity (i.e, H$^{-}$) affecting the formation of MgH and Mg lines is slightly different in stellar atmospheres. These observed inconsistencies seen directly from the comparison of line strengths, as in Figures \ref{comp_obs_spec_mgh}$-$\ref{comp_l26067_l33164mg}, are reflected in the [Mg/H] abundances (Table \ref{abu_helium}) derived from the atomic Mg and MgH lines when the standard stellar atmosphere with solar He/H$-$ratio is adopted. 
The relative variations in line strengths for MgH but not obviously for Mg lines (LEID 32149 and LEID 42042 in Figure \ref{comp_obs_spec_mgh} and Figure \ref{comp_obs_spec_mg}), and vice versa (LEID 26067 and LEID 33164 in Figure \ref{comp_l26067_l33164mgh} and Figure \ref{comp_l26067_l33164mg}) provide a direct confirmation of He-enhanced stellar photosphere of LEID 42042 (relative to LEID 32149) and LEID 26067 (relative to LEID 33164). This observational evidence acknowledge the reliability of abundance analysis technique and the He abundances derived for giants in this paper (Table \ref{abu_helium}).

\begin{table*}
{\fontsize{8}{8}\selectfont
\caption{Sensitivity of atomic Mg and MgH line abundances to uncertainties in stellar parameters exemplarily for a star LEID 42042 representative of red giants in this paper. Each number refer to the difference between the abundances obtained with and without varying each stellar parameter separately, while keeping the other parameters unchanged. Sensitivity of He abundance and mass fraction of helium ($Y$) corresponding to changes in [Mg/H] from MgH and Mg lines are listed.   }
\label{sensitivity_He}
\begin{tabular}{cccccc}   \hline
\multicolumn{1}{c}{Species} &\multicolumn{1}{c}{T$_{\rm eff}\pm$50} &\multicolumn{1}{c}{$\log\,g\pm$0.1} & \multicolumn{1}{c}{$\xi_{t}\pm$0.1} &
\multicolumn{1}{c}{[M/H]$\pm$0.1} & \multicolumn{1}{l}{ } \\ \cline{2-5} 

\multicolumn{1}{c}{(X)}&\multicolumn{1}{c}{$\delta_{X}$} &\multicolumn{1}{c}{$\delta_{X}$} & \multicolumn{1}{c}{$\delta_{X}$} &
\multicolumn{1}{c}{$\delta_{X}$}  &\multicolumn{1}{c}{$\sigma_{2}$} \\ \hline

 $[$Mg/H$]$$_{MgH}$ & $+0.10/-0.10$  & $-0.02/+0.03$   & $0.00/0.00$  & $+0.03/-0.03$   & 0.11  \\
  He   & $-0.03/+0.03$   & $+0.01/-0.01$   & $0.00/0.00$  & $-0.01/+0.01$   & 0.033 \\
  $Y$  & $-0.054/+0.054$ & $+0.017/-0.017$ & $0.00/0.00$  & $-0.017/+0.017$ & 0.059 \\
       &                 &                 &              &                 &         \\
 $[$Mg/H$]$$_{Mg}$ & $+0.03/-0.03$   & $0.01/-0.01$    & $-0.02/+0.04$   & $+0.01/0.00$ & 0.05  \\
   He          & $+0.01/-0.01$   & $+0.005/-0.005$ & $-0.01/+0.015$  & $+0.005/0.0$ & 0.02  \\
   $Y$         & $+0.017/-0.017$ & $+0.008/-0.008$ & $-0.017/+0.025$ & $+0.017/0.0$ & 0.029  \\
\hline
\end{tabular}
} \vspace{-0.4cm}
\end{table*}

\begin{figure}
\begin{center}
\includegraphics[trim=0.1cm 7.9cm 10.4cm 4.0cm, clip=true,height=0.3\textheight,width=0.47\textheight]{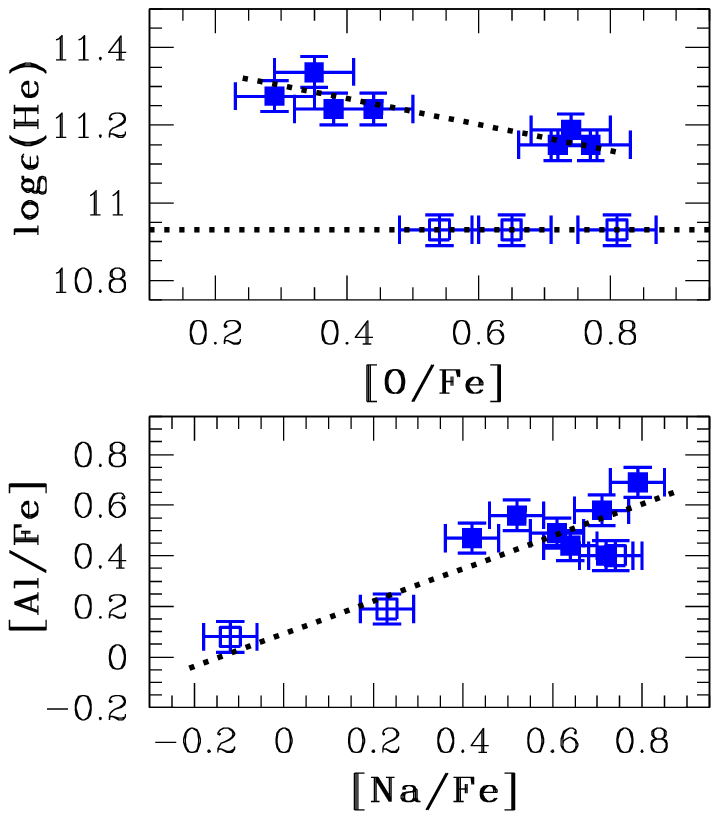} \vspace{-0.6cm}
\caption[]{The He$-$O anticorrelation (upper panel) and Na$-$Al correlation (Lower panel) observed for the present sample of Omega Cen red giant members. The He-enhanced and He-normal giants are represented by filled and open blue squares, respectively. The dotted line in each panel has the slope obtained from the least-squares fits to the data points. }
\label{he_light}
\end{center}
\end{figure}

Excluding the cluster non-members, the helium content for three of the member stars, namely LEID 33164, LEID 34029 and LEID 41476, is consistent with primordial value of $Y=$ 0.252 (Izotov et al. 2007). The highly enriched He content derived for the non-member giants may suggest either more evolved nature of these giants with stellar atmosphere replenished with helium via the dredge-up episodes and helium flash during RGB phase or the formation of these giants out of the ISM material enriched in He.
The average helium content of $Y=+$\,0.402$\pm$0.036 (or log\,$\epsilon$(He)$=$ 11.226$\pm$0.064 dex) for the remaining seven cluster members in Table \ref{abu_helium} exceeds $\Delta$$Y=+$0.15 the primordial value. Theoretical stellar evolution models predict after the first dredge-up that the $Y$ content of a star of metallicity $Z=$\,0.003 (roughly the metallicity of stars in this paper) increases by about 0.015 from its main sequence value (Sweigart \& Gross 1978; Sweigart 1997). Comparing with the first dredge-up value of $Y=$\,0.252$+$0.015\,$=$\,0.267, the $Y$ values for seven giants in Table \ref{abu_helium} are larger with a level of confidence between 1.5$\sigma$ (LEID 34180 and LEID 48319) to 2.8$\sigma$ (LEID 42042) of the internal error. If both internal and systematic errors in $Y$ are considered, this level of confidence is lower but marginally higher than 1$\sigma$ with a significance between 1.1$\sigma$ to 2.3$\sigma$. Therefore, the $Y$ values for seven cluster red giants presented here are significantly larger than the primordial value and those expected for a He-normal star after the first dredge-up. This result is robust within both internal and systematic errors affecting the $Y$ content of a star. To shift the mean $Y$ value to primordial value requires either the T$_{\rm eff}$ to be underestimated by about 200 K or the log~$g$ to be overestimated by about 0.7 dex. 
This result provides a direct confirmation of photometric predictions by Piotto et al. (2005) and Sollima et al. (2005) that stellar population highly  enhanced in He are present on the anomalous RGB-a of Omega Cen.

An indication of processed material in the atmospheres of these metal-rich and He-enhanced red giants of Omega Cen is provided from the inspection of elemental abundances for Na, Al and O. Figure \ref{he_light} shows He$-$O anticorrelation accompanied by Na$-$Al correlation with a clear enhancement of [Na/Fe] and [Al/Fe] in giant stars displaying enhanced helium content. Linear least-squares fits, following orthogonal regression method\footnote{It is the line that minimizes the sum of the square of the perpendicular distances between the data points and the line.} in Isobe et al. (1990), to the data points in Figure \ref{he_light} give slopes for log\,$\epsilon$(He) vs. [O/Fe] and [Al/Fe] vs. [Na/Fe] of $-$0.303$\pm$0.057 (R$=-$0.94) and $+$0.585$\pm$0.083 (R$=$\,0.91), where R is the Pearson product-moment correlation coefficient. These results signal contamination by products of high temperature H-burning process (Langer, Hoffman \& Sneden 1993, Prantzos, Charbonnel \& Iliadis 2007) including the CNO, NeNa and MgAl cycles in massive stars of previous stellar generations (Gratton et al. 2001).

The main product of H-burning is He with N being the by-product of CNO cycle at the expense of mainly C and a moderate amount of O, while the proton capture on Ne and Mg produces Na and Al, respectively. Examination of Figure \ref{he_light} shows, as anticipated, red giants with enhanced He content are also accompanied by enhancement of Na and Al abundances. Although the nitrogen abundance is not measured in this study, results from Simpson \& Cottrell (2013) generally confirm enhancement of N for all stars in common with this study. The mixing mechanism along the RGB is negligible, as the outer convective envelope in red giants will never reach the H-burning envelope where the light elements are synthesized during the evolution on the RGB (Gratton, Sneden \& Carretta 2004). The red giants of this paper show no dependence of helium abundances on spectroscopic luminosities (Table \ref{stellar_param}, thus, confirming evolutionary mixing along the RGB is negligible.

First acknowledgement of direct evidence for an enhancement of helium in Omega Cen is provided by Dupree et al. (2011) who measured Na, Al and Fe abundances and the He {\scs I} 10830 \AA\ line equivalent widths for a sample of 12 giants covering [Fe/H] range of $-$1.87 dex to $-$1.16 dex and a narrow range in effective temperature between 4560 K and 4770 K. They noted that five of the sample stars enhanced in helium are accompanied with enhancement of [Na/Fe] and [Al/Fe] with stars presenting no detected helium line have low values of [Na/Fe] and [Al/Fe]. Stars analysed in this paper are metal-rich with values of [Fe/H] in the range $-$0.74 dex to $-$1.30 dex. A majority of them have an effective temperature of about 4100 K and a surface gravity of about 1.3 dex with a cool limit at 3800 K and hot limit at 4350 K with a correlating range in surface gravity. Stars analysed in this paper show no trend of helium and other elemental abundances with either [Fe/H] or stellar parameters. Results of this paper confirm independently Dupree et al. prediction that the majority of red giants in Omega Cen with enhanced helium content are those exhibiting no correlation of helium abundance with [Fe/H] but a clear correlating trend of [Na/Fe] with [Al/Fe] abundances.
The sample of stars analysed in this paper and Dupree et al. covers much of the [Fe/H] range ($-$1.87 to $-$0.74 dex) sampled by various stellar populations in Omega Cen. These samples provide a direct confirmation that the helium enhanced stars present at all metallicities along all evolutionary stages on the CMD of Omega Cen.

The near-infrared He {\scs I} 10830 \AA\ line used in the study of Dupree et al. (2011) disappears in stars with effective temperature cooler than 4400 K (Strader, Dupree \& Smith 2015), irrespective of evolutionary state, from where the lines of MgH molecules start appearing. The spectroscopic chemical composition analysis of MgH and Mg lines, as shown in this paper, is a powerful probe of the presence of helium in such cool giants, whereas the lines of helium, He {\scs I} 10830 \AA\ and 5876 \AA, are useful informative diagnostic of He content in warm (4400 K $-$ 11,500 K) metal-poor stars of Omega Cen (e.g., SGB, HB and a few MS stars, Dupree et al. 2011; Villanova et al. 2012). As noted in the introduction, the helium content measured for stars hotter than 11,500 K is not representative of original surface He content (Grundahl et al. 1999; Marino et al. 2014). Therefore, a complete census of He-enhanced stellar population on the RGB, SGB and MS sequences of Omega Cen can be retrieved from the analysis of Mg, MgH and helium lines.

\begin{figure}
\begin{center}
\includegraphics[trim=0.1cm 6.5cm 6.3cm 4.0cm, clip=true,height=0.35\textheight,width=0.47\textheight]{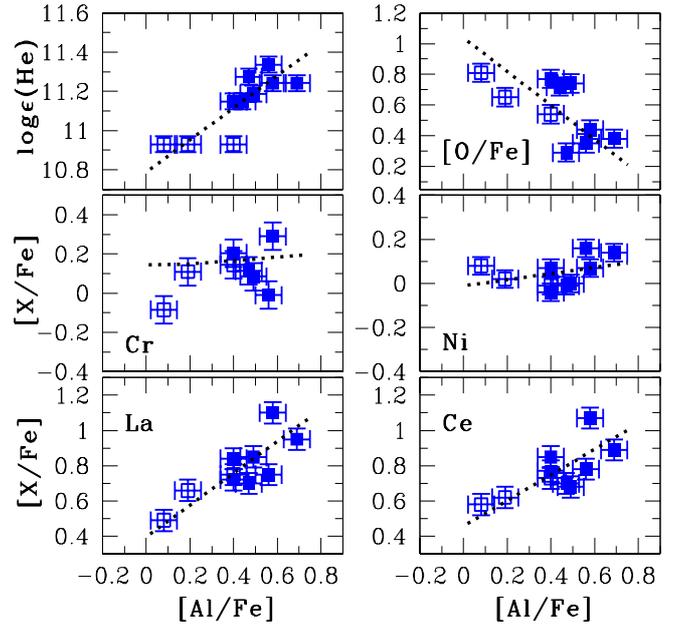} \vspace{-0.6cm}
\caption[]{The abundances of He, O, Cr, Ni, La and Ce plotted against the Al abundances for the He-enhanced (filled blue squares) and He-normal (open blue squares) cluster giants analysed in this paper. The dotted line in each panel has the slope obtained from the least-squares fit to the data points. No clear correlation is seen for [X/Fe] vs. [Al/Fe] for X$=$\,Cr and Ni.  }
\label{al_heavy}
\end{center}
\end{figure}

\subsection{Correlation of light and heavy elements}  \label{correlation_heavy_elements}
This section describes the dependence of Al abundances on He, O, iron-group and heavy elemental abundances to explore the possible sites of element production and eventual return via mass-loss and explosions to the ISM out of which the red giants of this study may have formed.

Variations from star-to-star in Al abundances are at the precision provided by abundance analyses of red giants uncorrelated with abundances of iron-group elements (Fe, Ni, and Cr) but anti-correlated with [O/Fe] and well correlated with the abundances of He and $s$-process elements (Ba, La, Ce, Sm). Figure \ref{al_heavy} shows abundances of He, O and two representative elements from iron-group (Cr and Ni) and $s$-process groups (La and Ce) plotted against Al abundances. 
Linear least-squares fit to the data points corresponds to a slope of $+$0.819$\pm$0.176 (R$=$\,0.90) for log\,$\epsilon$(He) vs. [Al/Fe], and slopes for [X/Fe] vs. [Al/Fe] of $-$1.102$\pm$0.363 (R$=-$0.82), $+$0.086$\pm$0.513 (R$=$\,0.18), $+$0.138$\pm$0.131 (R$=$\,0.57), $+$0.904$\pm$0.209 (R$=$\,0.90), $+$0.726$\pm$0.232 (R$=$\,0.85) for X$=$\,O, Cr, Ni, La and Ce, respectively.
The correlating trend of abundances between Al and $s$-process elements is an indication that the source of nucleosynthesis responsible for the Al enrichment is likely the principle site for the production of $s$-process elements.
In the Galactic chemical evolution context, the heavy elements in the range Y$-$Sm are primarily synthesized via the weak $s$-process (Y, Zr) occuring during He and C shell hydrostatic burning in massive stars ($M\gtrsim$ 4 $M_{\odot}$; timescale$\sim$30 Myr to 300 Myr) and the main $s$-process (Ba$-$Sm) from thermally-pulsing AGB stars ($M\gtrsim$ 1\,$-$\,3 $M_{\odot}$; timescale$\sim$ 1 Gyr). These massive AGB stars ($M\gtrsim$ 4 $M_{\odot}$) may synthesis large quantities of Al via the high temperature hydrogen burning during the MS phase (Busso, Gallino \& Wasserburg 1999; Burris et al. 2000; Ventura et al. 2001; Karakas \& Lattanzio 2003). The anticorrelation of Al with O, as in Figure \ref{al_heavy}, is a direct confirmation of the high-temperature hydrogen burning in massive stars. 

The red giants analysed in this paper showing enrichment in La$-$Sm are also enriched in Y and Zr abundances. This result for the heavy elements -- Y, Zr, La, Ba, La, Ce and Sm -- for the cluster members examined here supports D'Orazi et al. (2011)'s suggestion about the $s$-process enrichment in Omega Cen resulting from the ejecta of AGB stars with mass $>$\,3 M$_{\odot}$. The chemical composition analysis of heavy elements (Y, Zr, La, Ce, and Pb) in twelve red giants of Omega Cen by D'Orazi et al. reveals that the $s$-process element enrichment in Omega Cen extends over the whole n-capture domain, covering both the first-peak (Y and Zr) and second-peak (La and Ce) elements simultaneously, and possibly up to the third-peak element lead (Pb). This peculiar abundance pattern of $s$-process elements is expected to be realized via the main $s$-process operation in AGB stars with mass $>$\,3 M$_{\odot}$, unlike the 1$-$3 M$_{\odot}$ stars responsible for the production of second-peak n-capture elements in the Galaxy. While a quantitative estimate of the mass range requires modelling of AGB stars, results of this paper (Figure \ref{al_heavy}) and those presented in D'Orazi et al. suggest that the ejecta from the previous generation of metal-poor massive AGB stars ($>$\,3 M$_{\odot}$) may result large enhancements of Al and $s$-process elements via different nucleosynthetic processes.

The correlation of abundances between Al (may synthesize on the MS phase of massive AGB stars) and He suggest appreciable contribution of helium and Al to the interstellar medium (ISM) via stellar winds from AGB stars (3 $-$ 8 $M_{\odot}$) on a timescale of 30 Myr to 300 Myr (Renzini 2008). AGB stars in the mass range 3 $-$ 8 $M_{\odot}$ are expected to experience the second dredge-up (2DU) shortly before reaching the AGB (Becker \& Iben 1979) and the 3DU and hot-bottom burning (HBB) during AGB phase, leading to the ejection of stellar envelope highly enriched in helium and N and depleted in C and O abundances (Renzini \& Voli 1981; Renzini 2008). However, the total amount of fresh helium produced in AGB stars ranging from the primordial value of $Y=$\,0.25 for 3 $M_{\odot}$ to $Y=$\,0.36 for 8 $M_{\odot}$ with an average of $Y=$\,0.31 is insufficient to match the observed helium content of Omega Cen and red giants in this paper. Therefore, pollution from other sources of nucleosynthesis must accumulate in the interstellar medium before the formation of next generation of helium enhanced satrs, as observed in this paper. Proposed polluters to account for the helium enrichment in Omega Cen are AGB stars (D'Antona et al. 2002; D'Antona \& Caloi 2004), fast rotating massive stars (Decressin et al. 2007), and massive binaries exploding as supernovae (De Mink et al. 2009). The processed material from all these channels can potentially enrich the existing ISM with products of high temperature H-burning including He abundance of about $Y=$\,0.4 (Renzini 2008).

The correlation of Al (and He) with the $\alpha$-elements (Mg, Si, Ca, and Ti) is weak with a spread of $\pm$0.2 dex about the mean trend of [Al/Fe] with [X/Fe]. This is an indication that a fraction of Al (and He) observed in these giants may have synthesized in previous generations of massive stars $M\gtrsim$ 10 $M_{\odot}$; timescale$\sim$10 Myr) exploding as Type II supernovae which are responsible for the production of $\alpha$-elements. The metal-rich nature of these giants over the first generation of old, metal-poor rMS population of Omega Cen further suggests a sizeable contribution of iron to the ISM from the Type Ia SNe from white dwarfs exceeding the Chandrasekhar mass limit (1.4 $M_{\odot}$; timescale$\sim$100 Myr). As the lifetimes of the progenitors (Type Ia, II SN and AGB stars) responsible for the synthesis of elements are quite different and pollute the molecular clouds to differing degrees at different times, i.e., about 30 Myr to 1 Gyr for the $s$-process enrichment from AGB stars of 1 $-$ 8 $M_{\odot}$, 10 Myr to 100 Myr for the production of $\alpha$- and iron-group elements in massive stars ($>$\,8 $M_{\odot}$), it seems more likely that the ejecta from all these nucleosynthetic sources may have accumulated inside the cluster potential well before the formation of next generation of stars enriched in He, Na, Al and the $s$-process elements to the levels observed in red giants of this paper. These chemical enrichment timescales are consistent with relative age difference of about 1\,$-$\,2 Gyr found in Sollima et al. (2005) between the He-enhanced metal-rich population and the first generation of old metal-poor, He-normal (primordial value of He) rMS population of Omega Cen.

\section{Evolutionary connections}
The chemical and evolutionary structure of various sequences found on the CMD of Omega Cen is quite complex. Intracluster abundance variations seen among stellar populations in Omega Cen offer the opportunity for chemically tagging (Freeman \& Bland Hawthorn 2002; Lambert \& Reddy 2016) and establishing an evolutionary connection of red giants of this paper with cluster population. Before executing this exercise, the chemical and population structure of Omega Cen will be reviewed below.

There exists an extensive literature on photometric investigation of stars populating all evolutionary sequences on the CMD but little of the available data concerns chemical composition analysis.  Recent studies reporting photometry and low- to medium-resolution spectroscopic abundances ([Fe/H] and a few elements) of stars on the MS, SGB and RGB of Omega Cen include Piotto et al. (2005), Sollima et al. (2005), Villanova et al. (2007), Marino et al. (2011), Pancino et al. (2011), Johnson \& Pilachowski (2010), and Simpson \& Cottrell (2013) with several of these papers including compilations drawn from the literature to investigate the evolutionary connection of various populations in Omega Cen.

Based on the low-resolution (R$=$\,6400) spectroscopic metallicities of MS stars and isochrone fits to the CMD, Piotto et al. (2005) found that the bifurcated main sequence represents stars of [Fe/H]$=-$\,1.68 dex (spread of 0.2 dex) and $Y=$\,0.246 for the rMS and [Fe/H]$=-$\,1.37 dex with a range in $Y$ from 0.35 to 0.45 for the bMS. The ridge line of the bMS is best fitted with an isochrone of $Y=$\,0.38 and metallicity of $-$1.37 dex. 
Detailed wide field photometry and chemical composition analysis of elements C, N, Ca, Ti, Fe and Ba in subgiants observed at a resolution of $R=$\,6400 by Villanova et al. (2007) reveals four distinct SGB sequences on the CMD, named as SGB-A, SGB-B, SGB-C and SGB-D in order of increasing metallicity and decreasing subgiant branch magnitude, with corresponding metallicity peaks in the range [Fe/H]$=-$\,2.0 dex to $-$0.6 dex (see, Figure 5, Figure 15 and Table 4 in Villanova et al. 2007). Using the measured chemical abundances of RGB stars from analysis of medium-resolution spectra (R= 18,000), Johnson \& Pilachowski (2010) identified four distinct RGB populations denoted as, following Sollima et al. (2005), RGB: MP ([Fe/H]$\leq$$-$1.6), MInt1 ($-$1.6$<$[Fe/H]$<$$-$1.3), MInt2+3 ($-$1.3$<$[Fe/H]$<$$-$0.9), and an anomalous metal-rich sequence RGB-a ([Fe/H]$>$$-$0.9), where MP and MInt stands for metal-poor and metal-intermediate, respectively. 

In order to establish an evolutionary sequence of stars from the MS through SGB to RGB using chemical abundances, a suggestion made by Villanova et al. (2007) is that stars on the rMS feeds into the subgiant branches A, B, and C, while stars from the bMS move through SGB populations B and C to RGB. They identified that the SGB sequence SGB-D merges into the main-sequence, termed as MS-a (Bedin et al. 2004), which is blended with the rMS population. From the inspection of various CMDs, Villanova et al., as listed in their Table 4, identified the following evolutionary sequences: rMS $\rightarrow$ SGB-A $\rightarrow$ RGB-MP, bMS $\rightarrow$ SGB-B, C $\rightarrow$ RGB-MInt1 and RGB-MInt2+3, and MS-a $\rightarrow$ SGB-D $\rightarrow$ RGB-a.

\begin{figure}
\begin{center}
\includegraphics[trim=0.1cm 10.45cm 8.4cm 4.1cm, clip=true,height=0.25\textheight,width=0.47\textheight]{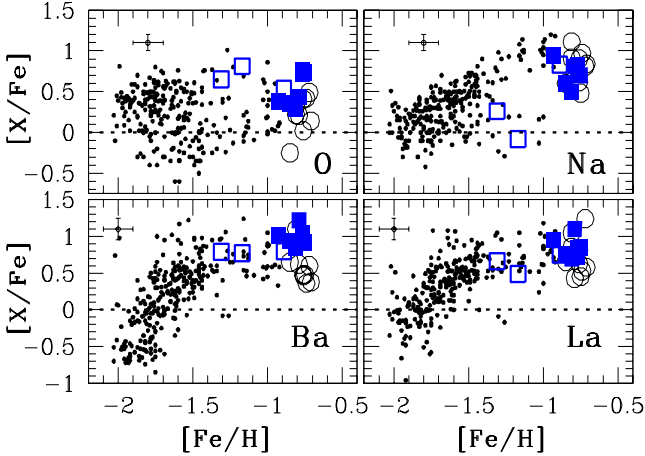} \vspace{-0.6cm}
\caption[]{ Comparison of [X/Fe] ratios for X$=$\,O, Na, Ba and La between Marino et al. (2011) sample of RGB-a giants (black filled circles for [Fe/H]$<$\,$-$0.95 dex and black open circles for [Fe/H]$>$\,$-$0.95 dex) and red giants in this work. He-enhanced and He-normal giants from this paper are represented by blue filled and open squares, respectively. }
\label{comp_marino2011}
\end{center}
\end{figure}

\begin{figure*}
\begin{center}
\includegraphics[trim=0.0cm 9.5cm 3.5cm 4.1cm, clip=true,height=0.25\textheight,width=0.85\textheight]{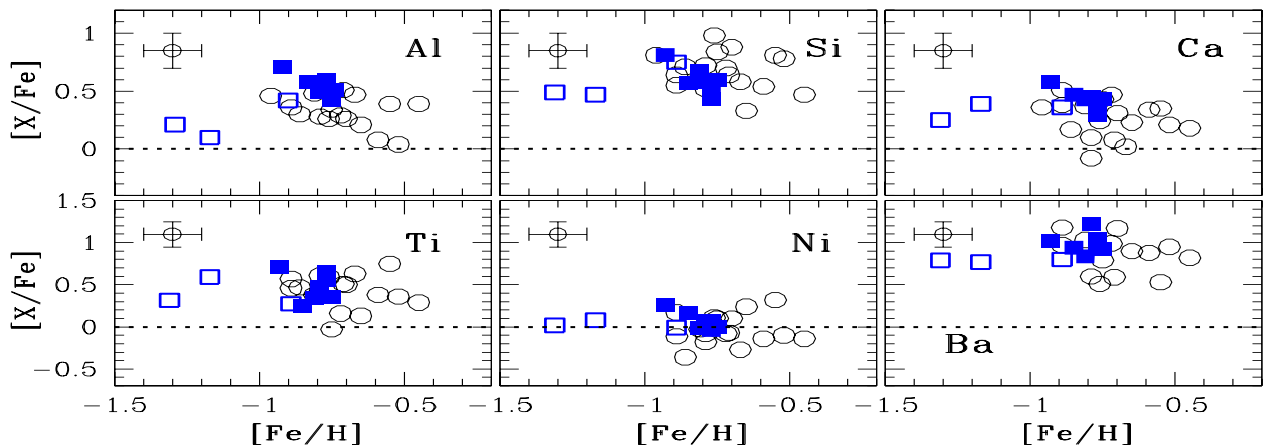} \vspace{-0.6cm}
\caption[]{ Comparison of [X/Fe] ratios for X$=$\,Al, Si, Ca, Ti, Ni, and Ba between Pancino et al. (2011) sample of subgiants on the SGB-a (black open circles) and red giants in this work (blue filled and open squares, respectively, represents the He-enhanced and He-normal giants).  }
\label{comp_pancino2011}
\end{center}
\end{figure*}

The evolutionary connection between the most metal-rich populations SGB-D (or SGB-a) and RGB-a is well established from studies of Pancino et al. (2002) and  Sollima et al. (2005). Pancino et al. (2002) measured from the high-resolution spectra (R$=$45,000) of three RGB-a stars a mean metallicity of [Fe/H]$=-$\,0.6$\pm$0.15 dex. Sollima et al. (2005) found that stars populating the most metal-rich subgiant branch (SGB-a; termed as SGB-D in Villanova et al. 2007) of Omega Cen has a mean metallicity of $-$0.6 dex fully compatible with that determined for stars on RGB-a.

The sample of Omega Cen giants analysed in this paper are metal-rich and span [Fe/H]$=$\,$-$0.7 to $-$1.3 dex with a mean value of $-$0.89$\pm$0.18 dex across the sample of 10 stars. They offer an excellent opportunity to explore the evolutionary connection of metal-rich populations in Omega Cen. Analogous to the idea of chemical tagging of open clusters (Freeman \& Bland Hawthorn 2002; Lambert \& Reddy 2016), identification of chemical similarities among the metal-rich giants of this paper and Omega Cen populations help us to unravel the evolutionary link of the majority of He-enhanced red giants in this paper.

Excluding the most metal-poor giants in this study (LEID 34029 and LEID 41476), the chemical abundances of other red giants of this paper will be discussed in comparison with recent chemical abundance results published for the metal-rich population of Omega Cen: RGB-a stars in Marino et al. (2011) and SGB-a stars in Pancino et al. (2011). 

A graphical comparison of [X/Fe] values for X$=$\,O, Na, Ba and La for the sample of Omega Cen giants analysed in this paper and RGB-a branch stars ([Fe/H]$>$\,$-$0.95 dex) of Marino et al. (2011) is provided in Figure \ref{comp_marino2011}. The typical measurement errors in [X/Fe] for individual stars in Marino et al. (2011) are of the order of 0.10 $-$ 0.15 dex in Ba and La, and are of 0.10 dex each for O and Na. These errors are comparable to typical errors in [X/Fe] found for giants in this paper. Inspection of the Figure \ref{comp_marino2011} shows for elements in common between the studies that the red giants in this paper generally share a common [X/Fe] with RGB-a branch stars across the sampled [Fe/H] range. Minor offsets in [X/Fe] may be related to the choice of spectral lines and methods used in the evaluation of stellar parameters (for example, the blended Ba{\scs II} line at 6141 \AA\ of Marino et al. vs. 5853 \AA\ barium line in this work, and photometric T$_{\rm eff}$ and log~$g$ of Marino et al. vs. spectroscopic parameters of this work). The difference in mean values of [Fe/H], [O/Fe], [Ba/Fe], and [La/Fe] between this work and Marino et al. of $-$0.01$\pm$0.10 dex, $+$0.32$\pm$0.23 dex, $-$0.15$\pm$0.21 dex, $+$0.35$\pm$0.28 dex, $+$0.06$\pm$0.28 dex, respectively, indicate almost complete agreement for [X/Fe] from all samples. The abundance results of Johnson \& Pilachowski (2010) for RGB giants in Omega Cen are generally confirmed by Marino et al. (2011). However, typically large offsets found in [X/Fe] between these studies is mainly attributed to the spectral quality, resolution, choice of spectral lines and analysis techniques adopted between studies.

A comparison of [X/Fe] values for X$=$ Al, Ca, Si, Ti, Ni, and Ba between SGB-a stars from Pancino et al. (2011) and red giants of this paper is provided in Figure \ref{comp_pancino2011}. The mean [X/Fe] of red giants in this study are in fair agreement with respect to mean values and scatter across the sampled [Fe/H]. Similarly, the [X/Fe] values for Ca ($+$0.48$\pm$0.03 dex), Ti ($+$0.44$\pm$0.06 dex) and Ba ($+$1.00$\pm$0.03 dex) for SGB-a stars in Villanova et al. (2007) are in good agreement with values found for red giants in this paper. Exception include [Fe/H] whose average value for SGB-a stars in Villanova et al. (2007) is less than about 0.5 dex. Villanova et al. (2007) results for chemical abundances of Ca, Ti, and Ba are generally confirmed by Pancino et al. (2011) with the sole exception of [Fe/H]. The average value of [Fe/H] from Villanova et al. (see, Figure 16 in Villanova et al. 2007) is lower by about 0.5 dex than that found for SGB-a stars in Pancino et al. study. This lower metallicity derived for the SGB-a stars in Villanova et al. is attributed to the selection of lines from blue spectral region (4400$-$4425 \AA) where the continuum placement due to heavy metal line-blanketing is quite uncertain (Pancino et al. 2011). In an independent low resolution ($R=$\,6500) study, Sollima et al. (2005) found from the analysis of infrared calcium triplet (8498 \AA, 8542 \AA, and 8662 \AA) an average metallicity of $-$0.6 dex for stars populating the SGB-a sequence. The chemical composition analysis of SGB-a stars by Sollima et al. (2005) and Pancino et al. (2011) ascertain that the SGB-a component has a mean metallicity of about $-$0.7$\pm$0.1 dex, comparable in magnitude to the average [Fe/H] of $-$0.79$\pm$0.06 dex derived for the He-enhanced red giants in this paper. In addition, the average [X/Fe] values for stars on SGB-a component agree very well with past abundance determinations of the RGB-a component (Pancino et al. 2002; Johnson \& Pilachowski 2010; Marino et al. 2011). 

The conclusion reached from the above discussion is that the He-enhanced red giants analysed in this paper are confirmed to be associated with the anomalous RGB-a population, which is identified as the bright-end continuation of the SGB-a sequence in all recent studies of photometry and spectroscopy (Sollima et al. 2005; Villanova et al. 2007; Bellini et al. 2010; Marino et al. 2011; Pancino et al. 2011).

Although He abundance is not available for individual metal-rich stars populating the SGB-a and RGB-a sequences, it was suggested based on photometric colours of stars by several authors that both SGB-a and RGB-a populations hosting the metal-rich stars of Omega Cen are enriched in helium with typical values between $Y=$\,0.35 and $Y=$\,0.45 (Norris 2004; Piotto et al. 2005; Sollima et al. 2005; Pancino et al. 2011). Such high values for the helium content are generally expected for stars on the bMS whose average metallicity exceeds 0.3 dex the metallicity of rMS population (Piotto et al. 2005; Villanova et al. 2007). However, the bMS population, on an average, a factor of 5 metal-poor than the SGB-a, RGB-a populations and red giants analysed in this paper (Piotto et al. 2005; Villanova et al. 2007).
Because of relatively metal-poor nature of bMS stars, the association of SGB-a and RGB-a population containing the most metal-rich ([Fe/H]$>$$-$0.9 dex) and the He-enhanced giants (this paper) with the bMS is discarded (i.e, [Fe/H] of about $-$1.37 dex for bMS vs $-$0.7 dex for SGB-a and RGB-a populations). The effect of helium content of the model atmospheres on the resulting abundance ratios including [Fe/H] is smaller than 0.15 dex (see also Pancino et al. 2011) to account for such a large metallicity difference (about $-$0.7 dex) between the bMS and SGB-a/RGB-a populations, confirming dissimilar evolution of the He-enhanced populations (bMS and MS-a) in Omega Cen. 
The He-enhancement between $Y=$\,0.36 and $Y=$\,0.46 and average [Fe/H] of $-$0.79$\pm$0.06 dex for the seven red giants in this paper (Table \ref{abu_helium}) is in fair agreement with the values of helium and metallicity inferred photometrically for the SGB-a and RGB-a populations (Sollima et al. 2005). Note that Villanova et al. (2007) didn't adopt a helium enhanced isochrone for fitting the SGB-a population on CMD. Note, however, based on the similarities in chemical abundances between SGB-a and RGB-a population, Pancino et al. (2011) suggest that SGB-a population is He-enhanced similarly to the RGB-a population.

Although metallicities and helium abundances for stars on the MS-a sequence are not available, because of the continuity of the He-enhanced, metal-rich isochrone from MS-a through SGB-a to RGB-a population (Sollima et al. 2005; Villanova et al. 2007), it is more likely that MS-a population is metal-rich and He-enhanced similarly to the SGB-a and RGB-a populations. Therefore, the metal-rich, He-enhanced giants analysed in this paper are the progeny of the MS-a population that evolved through SGB-a to RGB-a population on the CMD of Omega Cen. 
The chemical enrichment timescales, as discussed in Section \ref{correlation_heavy_elements}, and young age of MS-a, SGB-a and RGB-a populations by about 1\,$-$\,2 Gyr relative to bMS and rMS populations all support the evolutionary scenario advanced by Sollima et al. (2005) for the formation of metal-rich, He-enhanced population in Omega Cen i.e., the anomalous metal-rich population (MS-a, SGB-a, and RGB-a) of Omega Cen may have formed within 2 Gyr via either a rapid self-enrichment process, producing a large amount of helium ($Y=$\,0.4) and $s$-process elements ([$s$/Fe]$\simeq$\,$+$1.0 dex) or accretion of helium and other element enhanced stellar population evolved in a different environment by the Omega Cen. Based on the chemical similarity, the red giants of this paper provide the first confirmation that the RGB-a population is enhanced in helium including the $s$-process elements to the levels expected in a rapid self-enrichment process during the early stage of evolution of Omega Cen. However, the possibility exists that such an evolved population from a different environment may have absorbed in the early stage of Omega Cen (about 11.5 Gyr, Villanova et al. 2007) by the strong gravitational potential of the main component of Omega Cen hosting the bMS, rMS, SGB: A, B, C, and RGB: MP, MInt1, MInt2+3 population.

Detailed high-resolution spectroscopic abundance analysis for many elements including the helium content for several stars populating the MS, SGB and RGB including the MS-a and SGB-a sequences is essential for a thorough understanding of the origin of MS-a, SGB-a and RGB-a populations residing in Omega Cen. Such an excercise is left for a future paper. Nevertheless, the red giant stars of this paper presenting He-enhancement along with larger values for [Fe/H], Na, Al, and heavy elements provide a strong evolutionary connection with the MS-a, SGB-a and RGB-a populations than with other SGB and RGB populations including the bMS/rMS of Omega Cen.

\section{Conclusions}
This paper presents chemical composition analysis and the first measure of helium content for cool red giants populating the anomalous RGB-a component on the CMD of GC Omega Cen. 

The simultaneous abundance analysis of MgH and Mg lines based on the high-resolution and high S/N-ratio spectra adopting theoretical photospheric models with different mixtures of He/H$-$ratios, as shown in this paper, is the only powerful probe to evaluate the helium content of red giants cooler than 4400 K, where the helium line transitions (He {\scs I} 10830 \AA\ and 5876 \AA) are absent for a direct spectral line analysis (Dupree et al. 2011; Villanova et al. 2012). The derived He abundances reveal discovery of seven He-enhanced giants ($\Delta$$Y=+$0.15$\pm$0.04) among the RGB-a population, confirming the previously suggested photometric and isochrone based values of $Y=$\,0.35 $-$ 0.45 for RGB-a population (Sollima et al. 2005; Piotto et al. 2005). These seven giants represent the largest sample of giants, across the metallicity range ([Fe/H]$=-$\,2.0 to $-$0.6 dex) of Omega Cen, whose He content is measured from high-resolution spectroscopy. The presence of second generation(s) helium-enhanced stars on the RGB-a is strongly supported by the helium abundances measured from the high-resolution spectra of RGB-a giants. 

The present sample of giants confirm Dupree et al. (2011)'s claim that stars with He-enhanced atmospheres are apparently enriched in Na and Al indicative of high temperature H-burning including the CNO, NeNa and MgAl cycles that lead to He-production in previous stellar generations (Gratton et al. 2001). A straightforward He$-$O anticorrelation along with increasing trend of Al abundances with He, $\alpha$- and $s$-process element abundances supports that the He enrichment and other elements results from the mix of different proportions of ejecta from the fast rotating massive stars (Decressin et al. 2007), massive binaries exploding as supernovae (De Mink et al. 2009) and AGB stars of wide mass range into the ISM, out of which the He-enhanced red giants of this paper (RGB-a population) may have formed. The spread in metallicity and [X/Fe] values consistent with that of RGB-a and SGB-a populations signify a strong evolutionary link of red giants of this paper with RGB-a through SGB-a from a main-sequence stellar population (MS-a) enhanced in helium from previous stellar generations.

The confirmation of He-enhancement for the majority of stars on RGB-a (this paper), the evolutionary connection of RGB-a through SGB-a with MS-a than with bMS and the lack of large age difference among various evolutionary sequences of Omega Cen (Sollima et al. 2005) all suggest that the anomalous metal-rich population (MS-a, SGB-a, and RGB-a) of Omega Cen may have formed through either a rapid self-enrichment mechanism within 2 Gyr or accretion of a population evolved in a different environment (Sollima et al. 2005). Nevertheless, Omega Cen experienced a complex evolution with a puzzling chemical enrichment history which can be unravelled following chemical tagging where the tagging of stars will be established based on similarity in chemical composition including the He content. The realization of such an exercise requires homogeneous chemical composition analysis of many elements including the He content for several stars populating various MS, SGB and RGB sequences on the CMD of Omega Cen.

\vspace{0.1cm} 
{\bf Acknowledgements:} \\
I would like to thank Raffaele Gratton for carefully reading the manuscript and for giving a thoroughly helpful referee's report that helped improving the quality of the paper.
I thank Gajendra Pandey for helpful discussions. I would also like to thank David Lambert for helpful comments on an earlier version of the paper submitted to this journal.

Based on data products from observations made with ESO Telescopes at the La Silla Paranal Observatory under programs 67.D-0245(A), and 68.D-0332(A).
This paper includes data taken at The McDonald Observatory of The University of Texas at Austin. This publication makes use of data products from the Two Micron All Sky Survey, which is a joint project of the University of Massachusetts and the Infrared Processing and Analysis Center/California Institute of Technology, funded by the National Aeronautics and Space Administration (NASA) and the National Science Foundation (NSF).


\begin{thebibliography}{99}
\bibitem{} Alonso A., Arribas S., Mart\'{i}nez-Roger C., 1999, A\&AS, 140, 261
\bibitem{} Anderson J., 1997, PhD thesis, University of California, Berkeley
\bibitem{} Asplund M., Garc\'{i}a P\'{e}rez A. E., 2001, A\&A, 372, 601
\bibitem{} Asplund M., Grevesse N., Sauval A. J., Scott P., 2009, ARA\&A, 47, 481
\bibitem{} Becker S. A., Iben I. Jr., 1979, ApJ, 232, 831
\bibitem{} Bedin L. R., Piotto G., Anderson J., Cassisi S., King I. R., Momany Y., Carraro G., 2004, ApJ, 605, L125
\bibitem{} Bellini A., Bedin L. R., Piotto G., Milone A. P., Marino A. F., Villanova S., 2010, AJ, 140, 631
\bibitem{} Bergemann M., Collet R., Amarsi A. M., et al. 2017, ApJ, 847, 15
\bibitem{} B\"{o}hm-Vitense E., 1958, ZAp, 46, 1083
\bibitem{} Bono G., Iannicola G., Braga V. F., et al., 2019, ApJ, 870, 1152
\bibitem{} Busso M., Gallino R., Wasserburg G. J., 1999, ARA\&A, 37, 239
\bibitem{} Calamida A., Strampelli G., Rest A., Bono G., Ferraro I., Saha A., et al., 2017, AJ, 153, 175 
\bibitem{} Cannon R. D., Stobie R. S., 1973, MNRAS, 162, 207
\bibitem{} Castelli F., Kurucz R. L., 2003, IAU Symposium 210, Modelling of Stellar Atmospheres, Uppsala, Sweden, eds. N.E. Piskunov, W.W. Weiss, and D. F. Gray, 2003, ASP-S210
\bibitem{} Catelan M., 2009, Ap\&SS, 320, 261
\bibitem{} Charbonnel C., Brown J. A., Wallerstein G., 1998, A\&A, 332, 204
\bibitem{} Collet R., Asplund M., Trampedach R., 2007, A\&A, 469, 687
\bibitem{} Cutri R. M., Skrutskie M. F., van Dyk S., et al. 2003, The IRSA 2MASS All-Sky Point Source Catalog (Pasadena, CA: IPAC/California Institute of Technology)
\bibitem{} D'Antona F., Caloi V., 2004, ApJ, 611, 871
\bibitem{} D'Antona F., Caloi V., Montalb\'{a}n J., Ventura P., Gratton R., 2002, A\&A, 395, 69
\bibitem{} Decressin T., Meynet G., Charbonnel C., Prantzos N., Ekstr\"{o}m S., 2007, A\&A, 464, 1029 
\bibitem{} de Mink S. E., Cantiello M., Langer N., Pols O. R., Brott I., Yoon S. -Ch., 2009, A\&A, 497, 243
\bibitem{} D'Orazi V., Gratton R. G., Pancino E., et al. 2011, A\&A, 534, A29
\bibitem{} Dupree A. K., Avrett E. H., 2013, ApJ, 773, L28
\bibitem{} Dupree A. K., Strader J., Smith G. H., 2011, ApJ, 728, 155
\bibitem{} Freeman K., Bland-Hawthorn J., 2002, ARA\&A, 40, 4875
\bibitem{} Freeman K. C., Rodgers A. W., 1975, ApJ, 201, L71
\bibitem{} Fulbright J. P., McWilliam A., Rich R. M., 2007, ApJ, 661, 1152
\bibitem{} Gaia Collaboration, Brown A. G. A., Vallenari A., et al., 2018, A\&A, 616, 1G 
\bibitem{} Gaia Collaboration, Brown A. G. A., Vallenari A., et al., 2016, A\&A, 595, 2G
\bibitem{} Gallagher A. J., Caffau E., Bonifacio P., et al. 2016a, A\&A, 593, A4
\bibitem{} Gratton R. G., Bonifacio P., Bragaglia A., Carretta E., Castellani V., et al., 2001, A\&A, 369, 87
\bibitem{} Gratton R., Sneden C., Carretta E., 2004, ARA\&A, 42, 385
\bibitem{} Grundahl F., Catelan M., Landsman W.B., Stetson P.,B., Andersen M.I., 1999, ApJ, 524, 242
\bibitem{} Hinkle K. H., Wallace L., Ram R. S., Bernath P. F., Sneden C., Lucatello S., 2013, ApJS, 207, 26
\bibitem{} Isobe T., Feigelson E. D., Akritas M. G., Babu G. J., 1990, ApJ, 364, 104 
\bibitem{} Izotov Y. I., Thuan T. X., Stasi\'{n}ska G., 2007, ApJ, 662, 15
\bibitem{} Johnson C. I., Pilachowski C. A., 2010, ApJ, 722, 1373
\bibitem{} Johnson C. I., Pilachowski C. A., Simmerer J., Schwenk D., 2008, ApJ, 681, 1505
\bibitem{} Karakas A. I., Lattanzio J. C., 2003, PASA, 20, 279
\bibitem{} Lambert D. L., Reddy A. B. S., 2016, ApJ, 831, 202
\bibitem{} Langer G. E., Hoffman R., Sneden C., 1993, PASP, 105, 301
\bibitem{} Lattanzio J. C., 1986, ApJ, 311, 708
\bibitem{} Lee Y.-W., Joo J.-M., Sohn Y.-J., Rey S.-C., Lee H.-C., Walker A. R., 1999, Nature, 402, 55
\bibitem{} Lind K., Asplund M., Barklem P. S., Belyaev A. K., 2011, A\&A, 528, 103
\bibitem{} Lodders K., 2003, ApJ, 591, 1220  
\bibitem{} Maeckle R., Holweger H., Griffin R., Griffin R., 1975, A\&A, 38, 239
\bibitem{} Marino A. F., Milone A. P., Piotto G., Cassisi S., D'Antona F., Anderson J., et al., 2012, ApJ, 746, 14
\bibitem{} Marino A. F., Milone A. P., Piotto G., Villanova S., Gratton R., et al., 2011, ApJ, 731, 64
\bibitem{} Marino A. F., Milone A. P., Przybilla N., Bergemann M., Lind K., et al., 2014, MNRAS, 437, 1609
\bibitem{} Mayor M., Meylan G., Udry S., Duquennoy A., Andersen J., et al., 1997, AJ, 114, 1087
\bibitem{} Modigliani A., Mulas G., Porceddu I., Wolff B., Damiani F., Banse B. K., 2004, Msngr, 118, 8
\bibitem{} Modigliani A., Larsen J. M., 2012, FLAMES-UVES pipeline user manual, 14th edn. (ESO)
\bibitem{} Nordlander T., Lind K., 2017, A\&A, 607, A75
\bibitem{} Norris J. E., 2004, ApJ, 612, L25
\bibitem{} Norris J. E., Da Costa G. S., 1995, ApJ, 447, 680
\bibitem{} Osorio Y., Barklem P. S., Lind K., Belyaev A.K., Spielfiedel A., Guitou M., Feautrier N., 2015, A\&A, 579, A53
\bibitem{} Pace G., Recio-Blanco A., Piotto G., Momany Y., 2006, A\&A, 452, 493
\bibitem{} Pancino E., Ferraro F. R., Bellazzini M., Piotto G., Zoccali M., 2000, ApJ, 534, L83
\bibitem{} Pancino E., Mucciarelli A., Bonifacio P., Monaco L., Sbordone L., 2011, A\&A, 534, 53
\bibitem{} Pancino E., Pasquini L., Hill V., Ferraro F. R., Bellazzini M., 2002, ApJ, 568, 101
\bibitem{} Pasquini L., Avila G., Allaert E., Ballester P., Biereichel P., et al., 2000, SPIE, 4008, 129
\bibitem{} Pasquini L., Mauas P., Kaufl H.U., Cacciari C., 2011, A\&A, 531, 35
\bibitem{} Piotto G., Villanova S., Bedin L. R., Gratton R., Cassisi S., et al., 2005, ApJ, 621, 777
\bibitem{} Prantzos N., Charbonnel C., Iliadis C., 2007, A\&A, 470, 179 
\bibitem{} Ram\'{i}rez I., Allende Prieto C., 2011, ApJ, 743, 135
\bibitem{} Reddy A. B. S., Giridhar S., Lambert D. L., 2015, MNRAS, 450, 4301
\bibitem{} Reddy A. B. S., Giridhar S., Lambert D. L., 2012, MNRAS, 419, 1350
\bibitem{} Reddy A. B. S., Lambert D. L., 2019, MNRAS, 485, 3623R
\bibitem{} Reijns R. A., Seitzer P., Arnold R., Freeman K. C., Ingerson T., van den Bosch R. C. E., van de Ven G., de Zeeuw P. T., 2006, A\&A, 445, 503
\bibitem{} Renzini A., 2008, MNRAS, 391, 354
\bibitem{} Renzini A., Voli M., 1981, A\&A, 94, 175
\bibitem{} Simpson J. D., Cottrell P. L., 2013, MNRAS, 433, 1892 
\bibitem{} Simpson J. D., Cottrell P. L., Worley C. C., 2012, MNRAS, 427, 1153
\bibitem{} Sollima A., Pancino E., Ferraro F. R., Bellazzini M., Straniero O., Pasquini L., 2005, ApJ, 634, 332
\bibitem{} Sneden C., 1973, PhD Thesis, Univ. of Texas, Austin
\bibitem{} Stanford L. M., Da Costa G. S., Norris J. E., Cannon R. D., 2007, ApJ, 667, 911
\bibitem{} Strader J., Dupree A. K., Smith G. H., 2015, ApJ, 808, 124
\bibitem{} Sweigart A. V., 1997, ApJ, 474, L23 
\bibitem{} Sweigart A. V., Gross P. G., 1978, ApJS, 36, 405
\bibitem{} Thygesen A. O., Kirby E. N., Gallagher A. J., Ludwig H.-G., Caffau E., Bonifacio P., Sbordone L., 2017, ApJ, 843, 14
\bibitem{} Tull R. G., MacQueen P. J., Sneden C., Lambert D. L., 1995, PASP, 107, 251
\bibitem{} van de Ven G., van den Bosch R. C. E., Verolme E. K., de Zeeuw P. T., 2006, A\&A, 445, 513
\bibitem{} van Leeuwen F., Le Poole R.S., Reijns R.A., Freeman K.C., de Zeeuw P.T., 2000, A\&A, 360, 472
\bibitem{} Ventura P., D'Antona F., Mazzitelli I., Gratton R., 2001, ApJ, 550, L65
\bibitem{} Villanova S., Geisler D., Gratton R. G., Cassisi S., 2014, ApJ, 791, 107
\bibitem{} Villanova S., Geisler D., Piotto G., Gratton R., 2012, ApJ, 748, 62
\bibitem{} Villanova S., Piotto G., Gratton R.G., 2009, A\&A, 499, 755
\bibitem{} Villanova S., Piotto G., King I. R., Anderson J., Bedin L. R., et al., 2007, ApJ, 663, 296

\end{thebibliography}

\bsp	
\label{lastpage}
\end{document}